\documentclass[numbers,square,a4paper,fleqn]{cas-dc}

\usepackage[authoryear,longnamesfirst]{natbib}
\usepackage{upgreek}
\usepackage{mathrsfs}
\usepackage{amsthm}
\usepackage{amssymb}
\usepackage{amsfonts}
\usepackage{cancel}
\usepackage{graphicx}
\usepackage{color}
\usepackage{enumerate}
\usepackage{subcaption}
\usepackage{amsmath}
\usepackage{dcolumn}
\usepackage{bm}
\usepackage{hyperref}
\usepackage{float}
\usepackage{microtype}
\usepackage{tikz}
\usepackage[english]{babel}
\usepackage{multicol}
\usepackage{cuted}
\usepackage{changepage}

\def\tsc#1{\csdef{#1}{\textsc{\lowercase{#1}}\xspace}}
\tsc{WGM}
\tsc{QE}

\begin{document}
\let\WriteBookmarks\relax
\def\floatpagepagefraction{1}
\def\textpagefraction{.001}

\shorttitle{Shadows and photon spheres of black holes embedded in dark matter halo with a quintessential field}    

\shortauthors{Gong et al.}  

\title[mode = title]{Shadows and photon spheres of static black holes embedded in a Dehnen-(1,4,5/2)-type dark matter halo with a quintessential field}  



%

\author[1]{Hong-Er Gong}[orcid=0009-0009-3469-2379]
\ead{gonghonger@yeah.net}
\author[1]{Junlin Qin}[orcid=0009-0007-9764-2235]
\ead{junlin.qin@st.gxu.edu.cn}
\author[1]{Yusen Wang}[orcid=0009-0003-6494-3104]
\author[1]{Bofeng Wu}[orcid=0000-0001-7330-6455]
\cormark[1]
\ead{wubofeng@gxu.edu.cn}
\author[1]{Zhan-Feng Mai}[orcid=0000-0003-2724-0226]
\cormark[1]
\ead{zf1102@gxu.edu.cn}
\author[2,3,4]{Xiao Zhang}[orcid=0000-0001-5893-7523]
\cormark[1]
\ead{xzhang@amss.ac.cn}
\author[1]{Enwei Liang}[orcid=0000-0002-7044-733X]
\ead{lew@gxu.edu.cn}



\affiliation[1]{organization={Guangxi Key Laboratory for Relativistic Astrophysics},
            addressline={School of Physical Science and Technology, Guangxi University}, 
            city={Nanning},
            postcode={530004}, 
            country={China}}
\affiliation[2]{organization={Guangxi Center for Mathematical Research},
            addressline={Guangxi University}, 
            city={Nanning},
            postcode={530004}, 
            country={China}}
\affiliation[3]{organization={Academy of Mathematics and Systems Science},
            addressline={Chinese Academy of Sciences}, 
            city={Beijing},
            postcode={100190}, 
            country={China}}
\affiliation[4]{organization={School of Mathematical Sciences},
            addressline={University of Chinese Academy of Sciences}, 
            city={Beijing},
            postcode={100049}, 
            country={China}}
\cortext[1]{Corresponding author}

\begin{abstract}
This paper investigates the appearance characteristics of static black holes embedded in Dehnen-(1,4,5/2)-type dark matter halos with a quintessential field, focusing on how the dark matter halo and dark energy affect the black hole images. We first derive the event horizon radius and the photon effective potential of the black hole, and then calculate critical quantities such as the critical photon sphere radius and critical impact parameter under different parameter sets. Trajectories of photons are subsequently plotted. The study reveals that as the parameters of the dark matter halo (the central density of the dark matter halo $\rho_s$ and the scale radius of the central halo $r_s$) and the quintessential field (the normalization factor $c$ and the equation of state parameter of dark energy $w_q$) increase, the aforementioned physical quantities generally exhibit an increasing trend. Based on the derived general expressions for the redshift factor and integrated intensity, we further explore the optical effects of the spherical accretion and the thin-disk accretion models. The results indicate that dark energy exerts an influence on the black hole shadow that is strongly dependent on the observer's position, whereas the influence exerted by dark matter exhibits no such conspicuous dependence. Furthermore, dark matter and dark energy have distinct effects on both the intensity and the radius of the black hole shadow. In particular, the intensity exhibits a greater sensitivity to dark energy, whereas the radius is more responsive to dark matter. This distinction offers a potential observational criterion for identifying, through black hole images, whether the dominant interacting component near the black hole is dark matter or dark energy, and provides an important basis for constraining the equation-of-state parameter $w_q$.
\end{abstract}





\maketitle

\section{Introduction}\label{sub1}
Albert Einstein’s General Relativity (GR), one of the most successful theories of gravity, accurately describes gravitational interactions and predicts the existence of black holes (BHs)~\citep{r1,r2,r3}. In recent years, the Laser Interferometer Gravitational Wave Observatory (LIGO) and Virgo collaborations have detected gravitational waves from binary BH mergers~\citep{r4,r5,r6,r7}, while the Event Horizon Telescope (EHT) has captured images of the supermassive BHs M87* and Sgr A*~\citep{r8,r9,r10,r11,r12,r13,r14,r15,r16,r17,r18,r19,r20,r21,r22,r23,r24}. These breakthroughs provide direct and indirect confirmation of BHs, strongly supporting GR.

Despite these triumphs in strong-field gravity, GR faces a profound challenge on cosmological scales—it cannot account for the dominant constituents of the universe, dark matter (DM) and dark energy (DE). Observations such as the cosmic microwave background radiation indicate that the universe consists of approximately 26.8\% DM and 68.5\% DE~\citep{r59,r60}.
Although DM has not been directly detected,  its existence is robustly supported by a wealth of indirect astrophysical evidence~\citep{r61,r62,r63,r64,r65,r66}. Depending on the model parameters, the DM density distribution can be described by various profiles~\citep{r67,r68,r69,r70}. 

In 1993, Dehnen proposed a generalized family of spherically symmetric density profiles parameterized by the inner slope $\gamma$~\citep{r71}.
Compared with the Navarro–Frenk–White (NFW) model characterized by its distinctive cuspy inner density profile, the Dehnen model provides a more flexible framework for analyzing DM density distributions under different scenarios, and can even encompass multiple classical profiles as special cases. For instance, $\gamma = 1$ corresponds to the Hernquist profile, $\gamma = 2$ corresponds to the Jaffe profile, and $\gamma = 0$ represents a cored profile structure. Owing to its flexibility, the Dehnen model has been widely applied in the fields of galactic dynamics and BH astrophysics~\citep{d1,d2,r74,r75,r76,r77,r78}.
The Dehnen-$(\alpha, \beta, \gamma)$ profile can be expressed as
\begin{equation*}
    \rho = \rho_s \left(\frac{r}{r_s}\right)^{-\gamma} \left[\left(\frac{r}{r_s}\right)^{\alpha} + 1\right]^{\frac{\gamma-\beta}{\alpha}},    
\end{equation*}
where $\rho_s$ denotes the central density of the DM halo and $r_s$ denotes the scale radius of the central halo.
The Dehnen-(1, 4, 5/2) DM halo profile adopted in this work is selected mainly on the basis of three physical considerations:
First, in the Dehnen profile model, different values of $\gamma$ correspond to different DM density distributions, with a typical valid range of $0<\gamma<3$~\citep{r71}. The parameter set $(1,4,5/2)$ employed in this work can reasonably describe the DM environment surrounding supermassive BHs at galactic centers, and thus bears clear astrophysical significance.
Second, for a static BH embedded in a DM halo with this parameter set, there exists an exact analytical metric solution to the Einstein field equations, with the metric function given by an explicit elementary expression. Unlike DM halo models that require numerical approximation to construct the BH spacetime, the analytical metric ensures that spacetime properties are continuously differentiable to high orders throughout the entire region outside the event horizon. This effectively avoids interpolation errors and numerical instability in strong-field calculations associated with numerical metrics, thereby providing a high-precision spacetime background for the ray-tracing and shadow imaging computations in this work.
Third, the spacetime is rigorously physically self-consistent. As verified in Reference~\citep{e1}, this spacetime satisfies all four classical energy conditions — weak, null, strong, and dominant — in all physical regions outside the event horizon. Curvature invariants such as the Ricci scalar and the Kretschmann scalar remain finite everywhere, with a spacetime singularity present only at the center of the BH. This guarantees that all physical results obtained in this work based on this spacetime comply with fundamental relativistic physical constraints.

Observations of Type Ia supernovae have confirmed the accelerated expansion of the universe~\citep{r79,r80,r81}. Within the framework of GR, this acceleration requires a cosmic component with negative pressure, i.e., DE. A prominent candidate for DE is quintessence, a dynamical scalar field, characterized by an equation of state parameter $w\equiv p/\rho$, where $p$ and $\rho$ denote the pressure and energy density of the quintessential field (QF), respectively. When $w<-1/3$, quintessence can drive the observed cosmic acceleration~\citep{r82,r83}. In the context of BH physics, Kiselev derived a static BH solution surrounded by a QF~\citep{r84}, providing a foundational framework for subsequent studies in this area, including the present work.

Light rays near a BH undergo deflection due to gravitational lensing. Photons with an impact parameter less than the critical value are captured, creating a distinct dark region on the observer's sky against the surrounding emission. This dark region, known as the BH shadow, is a direct projection of the photon sphere and is thus a powerful probe of the underlying spacetime geometry~\citep{r25,r26,r27,r28}. The groundbreaking images of M87* and Sgr A* from the EHT have established shadow observations as a precise new test of strong-field gravity ~\citep{r29,r30,r31,r32,r33,r34,r35,r36,r37,r38,r39}. Therefore, analyzing the features of BH shadows provides a unique avenue to investigate the effects of DM and DE in BH imaging.
The existence of DM and DE implies that BHs are inevitably embedded in these components, which alter the surrounding spacetime geometry and thereby distort their optical appearance. Consequently, analyzing BH images offers a promising avenue not only to probe the properties of DM and DE but also to understand their gravitational influence on BHs~\citep{r78, r85}.
To fully realize the potential of this approach, the coupled effects of DM and DE must be considered together. However, most studies to date have considered them in isolation, even though in realistic cosmic environments, BHs are simultaneously affected by both as they jointly shape the spacetime metric~\citep{r86,r87}. As a result, research on BH shadows and images within such composite systems remains limited. Therefore, a systematic investigation of their coupled effects is required.
To address this research gap, this paper systematically studies, based on the metric proposed in reference~\citep{r87}, the optical characteristics of static BHs embedded in a Dehnen-(1,4,5/2)-type DM halo with a QF.

The first core objective of this work is to systematically characterize the BH image features under the spherical accretion framework. We begin with the spherical accretion model, an idealized configuration of spherically symmetric matter that has been widely adopted in recent theoretical studies of BH imaging~\citep{r48,r49,r50,r51,r52}. Within this model, we examine in detail two physically distinct limiting cases: (i) static spherical accretion, in which the accreting fluid has negligible radial velocity, and (ii) infalling spherical accretion, where the flow is in free fall toward the BH with a substantial inward radial velocity. In the infalling case, the inward motion induces significant Doppler redshift, which directly suppress the observed radiation intensity. As a result, the shadow under static accretion appears consistently brighter than that under infalling accretion—a direct consequence of the Doppler redshift in the infalling scenario.

With the two accretion scenarios, our analysis reveals a series of distinct and observationally relevant imprints of the DM halo and QF on BH shadow properties within the spherical accretion framework. First, we find that the shadow radius for both static and infalling spherical accretion is strictly equal to the critical impact parameter. This result confirms that the shadow size is a purely geometric probe, determined entirely by the background spacetime geometry and independent of the details of the accretion flow. Regarding the radiative properties of the shadow, the DM and QF parameters exert opposing influences: increasing the DM parameters (central density $\rho_s$ or scale radius $r_s$) reduces the observed shadow intensity, whereas increasing the QF parameters (normalization factor $c$ or the magnitude of the equation-of-state parameter $|w_q|$) enhances it. We further uncover an interdependence among the parameters within each sector: for the DM halo, a larger $\rho_s$ amplifies the sensitivity of the intensity to changes in $r_s$, and vice versa; an identical mutual enhancement of sensitivity exists between the QF parameters $c$ and $|w_q|$. Finally, we highlight a critical and observationally distinguishable difference between the DM and QF effects: unlike the DM-induced intensity modulation, which shows negligible dependence on the observer’s location, the influence of the QF on the shadow intensity depends strongly on the observer’s distance from the BH. Specifically, for fixed increases in the parameters $c$ and $|w_q|$, a more distant observer measures a higher shadow intensity.

The second core objective of this work is to systematically investigate the optical appearance and shadow properties of the BH under the canonical face-on thin-disk accretion framework, rigorously formulated and analyzed in reference~\citep{r54}. This framework, featuring an optically and geometrically thin accretion disk viewed from a face-on orientation, provides a physically realistic and analytically tractable description of the luminous accreting matter around most astrophysical supermassive and stellar-mass BHs. The radiation emitted from such disk structures serves as the primary observational source for modern BH imaging experiments, rendering this thin-disk framework the fundamental baseline for interpreting the observed signatures of horizon-scale BH emission. The pioneering work in reference~\citep{r54} resolved a long-standing ambiguity: it distinguished the heuristic “BH shadow” (defined for a uniform background) from the actual dark region produced by accretion emission. Moreover, it established a rigorous analytic framework to decompose the observed image into direct emission, lensing ring, and photon ring components. Since this landmark study, the thin-disk accretion scenario, particularly the canonical face-on thin-disk model as the standard benchmark configuration, has been widely adopted, extended, and tested in a wealth of subsequent theoretical developments and observational data interpretation studies~\citep{r55,r56,r57,r58}. In this part of our work, we will strictly follow the standard analytic methodology and model prescriptions established in this classic literature, to perform a detailed and self-consistent exploration of BH optical appearance and shadow characteristics in the face-on thin-disk accretion regime.

We systematically characterize the geometric structure and imaging morphology of BH images within the thin‑disk scenario, which exhibits distinct features compared to the spherical accretion case. Three distinct thin accretion disk models are investigated, with their inner radiation boundaries set to the innermost stable circular orbit (ISCO), the photon sphere, and the event horizon, respectively. The results indicate that the choice of the accretion disk’s inner boundary directly modulates the apparent shadow radius of the BH: the model with the ISCO as the inner boundary yields the largest shadow, followed by the model with the photon sphere as the inner boundary, while the model with the event horizon as the inner boundary produces the smallest shadow. This boundary-dependence contrasts sharply with the spherical accretion case, where the shadow radius is strictly determined by spacetime geometry and independent of accretion details. Furthermore, we demonstrate that the inner boundary of the disk dictates the separability of emission components in the BH image. When the ISCO serves as the inner boundary, the direct emission, lensed ring, and photon ring are clearly distinct, manifesting as three separate peaks in the intensity profile. For the photon sphere boundary, the lensed and photon rings superimpose directly onto the direct emission, resulting in only two observable peaks. Conversely, when the event horizon defines the emission boundary, although the ring components overlap with the direct emission, they still give rise to two distinguishable peaks corresponding to the photon ring and lensed ring, respectively. In addition, we find that increasing values of both the DM parameters ($\rho_s$, $r_s$) and QF parameters ($c$, $|w_q|$) lead to a spatial broadening of the lensed ring and photon ring regions, with a similar mutual enhancement of parameter sensitivity as in the spherical accretion case: a larger value of $\rho_s$ (or $r_s$) amplifies the sensitivity of the ring regions to variations in $r_s$ (or $\rho_s$), and vice versa for the QF parameters $c$ and $|w_q|$. Notably, the radial positions of the lensed and photon rings remain robust against changes in the observer's inclination, providing a stable geometric probe of the underlying spacetime.

Concerning the radiative properties of BH images for the thin-disk scenario, our analysis demonstrates both consistencies and key differences relative to the spherical accretion case. First, we find that the integrated intensity is universally dominated by direct disk emission across all three thin-disk models, while the contributions from the lensed and photon rings remain negligible due to their narrow spatial extent. Second, the opposite effects of DM and QF on the radiation intensity are fully consistent with the spherical accretion results: increasing the DM parameters $\rho_s$ and $r_s$ suppresses the integrated intensity, whereas increasing the QF parameters $c$ and $|w_q|$ significantly enhances it. Third, we highlight a critical observer-dependent effect analogous to that observed in spherical accretion case: the influence of the DM parameters on the integrated intensity is nearly independent of the observer’s radial position, whereas the effect of the QF parameter $w_q$ on the intensity exhibits a strong dependence on the observer’s distance — for a fixed increase in $|w_q|$, a more distant observer measures a higher integrated intensity. In contrast, the influence of the QF normalization parameter $c$ on the intensity remains weakly dependent on the observer's position, particularly at small values of $|w_q|$.
These distinct, decoupled signatures of DM and QF on the geometric structure (shadow size, ring broadening) and radiative intensity of BH images under different accretion models provide a robust observational criterion to discern whether the gravitational environment of a BH is dominated by DM or DE, and offer an important independent basis for constraining the DE equation-of-state parameter $w_q=p_q/\rho_q$ using high-resolution BH imaging data.

The remainder of this paper is organized as follows. Section~\ref{sub2} reviews the metric of the static BH embedded in a Dehnen-(1,4,5/2)-type DM halo with a QF, and discusses the variation of the event horizon. Section~\ref{sub3} applies the geodesic equations and effective potential to study light deflection and the critical impact parameter for such a BH, and plots the light trajectories. Section~\ref{sub4} explores a spherically symmetric accretion model, presents the intensity profiles and appearance of the BH, and generates its image. Section~\ref{sub5} provides an in-depth analysis of the observational effects of an optically and geometrically thin accretion disk around the BH. Section~\ref{sub6} summarizes and discusses the findings of this study. Throughout this work, geometrical units are adopted, with the speed of light and Newton’s gravitational constant set to unity.

\section{Metric for a static BH in a Dehnen-(1,4,5/2)-Type DM halo with a QF}\label{sub2}
We consider a static BH model within a Dehnen-(1, 4, 5/2)-type DM halo with a QF. The corresponding metric line element is given by~\citep{r87}
\begin{equation}
    \mathrm{d}s^2=-f(r)\mathrm{d}t^2+\dfrac{1}{f(r)}\mathrm{d}r^2+r^2(\mathrm{d}\theta^2+\sin^2\theta\mathrm{d}\varphi^2),\label{ds}
\end{equation}
where the metric function $f(r)$ is
\begin{equation}
    f(r)=1-\dfrac{2M}{r}-32\pi\rho_sr_s^2\sqrt{\dfrac{r+r_s}{r}}-\dfrac{c}{r^{3w_q+1}}.\label{f}
\end{equation}
Here $M$ denotes the BH mass, $\rho_s$ the central density of the DM halo, and $r_s$ the scale radius of the central halo. The parameter $c$ is a normalization factor associated with the QF DE density, and $w_q$ is the proportionality coefficient in the DE equation of state $p_q=w_q\rho_q$, i.e., the ratio of the QF pressure $p_q$ to its density $\rho_q$~\citep{r84}. According to reference~\citep{r84}, any QF with $w_q < -1/3$ can act as DE, and in this case, the relation between $c$ and $\rho_q$ is
\begin{equation}
    \rho_q=-\dfrac{c}{2}\dfrac{3w_q}{r^{3(1+w_q)}},
\end{equation}
which shows that $c$ directly reflects $\rho_q$.

\begin{figure*}[tp]
    \centering
            \includegraphics[width=0.45\textwidth]{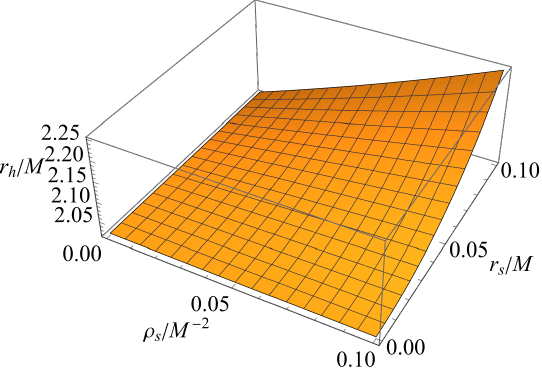}
            \hspace{0.5cm}
            \includegraphics[width=0.45\textwidth]{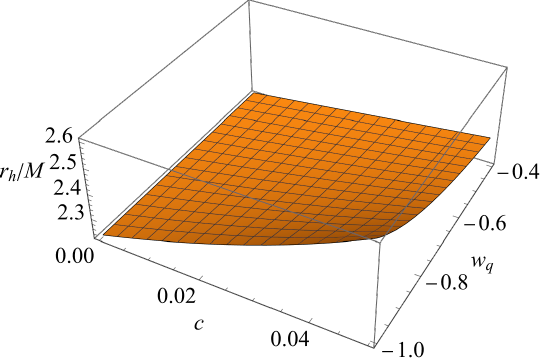}
    \caption{
    Left panel: Dependence of the event horizon radius $r_h/M$ on the DM parameters $\rho_s/M^{-2}$ and $r_s/M$, for fixed QF parameters ($c=0.01$ and $w_q=-0.4$).
    Right panel: Dependence of the event horizon radius $r_h/M$ on the QF parameters $c$ and $w_q$, for fixed DM parameters ($\rho_s/M^{-2}=0.1$ and $r_s/M=0.1$).}
    \label{rhdm}
\end{figure*}

For BH with the metric~\eqref{ds}, the event horizon is derived from $f(r_h)=0$, which means that $r_h$, as the event horizon radius, satisfies
\begin{equation}
    1-\dfrac{2M}{r_h}-32\pi\rho_sr_s^2\sqrt{\dfrac{r_h+r_s}{r_h}}-\dfrac{c}{r_h^{3w_q+1}}=0.\label{rhoo}
\end{equation}
We solved eq.~\eqref{rhoo} using numerical methods, and the results are plotted in figure~\ref{rhdm}. 
The left panel of figure~\ref{rhdm} shows the variation of the BH event horizon radius $r_h/M$ with $\rho_s/M^{-2}$ and $r_s/M$ for a fixed QF ($c=0.01$, $w_q=-0.4$). Here, $r_h/M$ increases with increasing $\rho_s/M^{-2}$ and $r_s/M$. The right panel shows how $r_h/M$ depends on $c$ and $w_q$ for a fixed Dehnen-(1,4,5/2) DM halo ($\rho_s/M^{-2}=r_s/M=0.1$). It can be seen that $r_h/M$ increases with $c$ but decreases with $w_q$.
Note that when the QF is absent, i.e., $c=0$, eq.~\eqref{f} reduces to the metric function of a BH embedded solely in a DM halo~\citep{e1}
\begin{equation}
    f(r)=1-\dfrac{2M}{r}-32\pi\rho_sr_s^2\sqrt{\dfrac{r+r_s}{r}},\label{dh}
\end{equation}
and in this case, eq.~\eqref{rhoo} has an analytical solution
\begin{equation}
    r_{h\mathrm{DM}}=\dfrac{2M+g^2r_s/2+g\sqrt{g^2r_s^2/4+2M(r_s+2M)}}{1-g^2},
\end{equation}
where $g=32\pi\rho_sr_s^2$. This shows that the event horizon exists only when $\rho_sr_s^2<1/32\pi$.
Our results, shown in the left panel of figure~\ref{rhdm}, are consistent with reference~\citep{r78} for the pure DM halo case: increasing $\rho_s$ and $r_s$ enlarges the event horizon, and this qualitative behavior persists even with the inclusion of a QF.
Similarly, when DM is not considered, i.e., $\rho_s=0$, eq.~\eqref{f} reduces to the metric function of a Kiselev BH~\citep{r84}
\begin{equation}
    f(r)=1-\dfrac{2M}{r}-\dfrac{c}{r^{3w_q+1}}.\label{de}
\end{equation}

The right panel of figure~\ref{rhdm} shows that this phenomenon also exists in the QF parameters: even in the presence of a Dehnen-(1, 4, 5/2)-type DM halo, as $c$ and $|w_q|$ increase, the event horizon expands, which is consistent with the qualitative results of the pure QF case~\citep{r85}.
Further, for $w_q=-1$, eq.~\eqref{de} reduces to the metric function of a Schwarzschild–de Sitter BH
\begin{equation}
    f(r)=1-\dfrac{2M}{r}-cr^2,\label{2.8}
\end{equation}
where $c$ is related to the cosmological constant $\mathnormal{\Lambda}$ by $c=\mathnormal{\Lambda}/3$. This model contains both an event horizon $r_{h\mathrm{DE}}$ and a cosmological horizon $r_{c\mathrm{DE}}$. When $w_q=-2/3$, the expressions for the event horizon and the cosmological horizon are respectively
\begin{equation}
    r_{h\mathrm{DE}}=\dfrac{1-\sqrt{1-8Mc}}{2c},\qquad r_{c\mathrm{DE}}=\dfrac{1+\sqrt{1-8Mc}}{2c}.\label{2.9}
\end{equation}
The above results demonstrate that the BH enters the extremal state when $r_{h\mathrm{DE}}=r_{c\mathrm{DE}}$, which implies that the two horizons exist only when $8Mc<1$. In this work, we exclude BHs in the extremal state, such that the observer under consideration lies between the event horizon $r_h$ and the cosmological horizon $r_c$.
When $w_q=-1/3$, the metric function in eq.~\eqref{de} degenerates to that of a Schwarzschild BH with only an event horizon (up to a constant scale factor).

From the explicit form of the metric function given in eq.~\eqref{2.8}, we can directly see that for the spacetime described by metric~\eqref{ds}, when the QF parameter $(c,w_q)$ takes specific values (e.g., $c<1/8M$, $w_q=-2/3$), the spacetime not only possesses the event horizon $r_h$ discussed above, but also develops an additional cosmological horizon $r_c$ outside the event horizon. In this case, the spacetime deviates from asymptotic flatness, and the region between the two horizons is defined as the domain of outer communication, where any two observers can exchange signals without being blocked by the horizons~\citep{r97, r98}. Therefore, the radial coordinate of the observer's position $r_\mathrm{obs}$ must lie between the event horizon and the cosmological horizon, i.e.,
\begin{equation}
    r_h<r_\mathrm{obs}<r_c.
\end{equation}
\section{Impact parameter, effective potential, and light rays\label{sub3}}
To obtain the image properties of a BH, we need to analyze the light trajectories. For this purpose, we will study the null geodesics of photons emitted by the accretion flow. Without loss of generality, we confine the null geodesics to the equatorial plane, i.e., set $\theta=\pi/2$. For the spherically symmetric metric described in eq.~\eqref{ds}, there exist two Killing vector fields $\xi^\mu_1$ and $\xi^\mu_2$, whose explicit expressions are
\begin{equation}
    \xi^\mu_1=\updelta_t^\mu,\qquad\xi_2^\mu=\updelta_\varphi^\mu.
\end{equation}
Let the photon’s 4‑momentum be $K^\mu=\mathrm{d}x^\mu/\mathrm{d}\lambda$, where $\lambda$ is the affine parameter along the photon worldline. Then, on a null geodesic, there are two conserved quantities: the photon energy $E$ and angular momentum $L$, given respectively by
\begin{equation}
\begin{split}
    E=&-g_{\mu\nu}\xi_1^\mu\dfrac{\mathrm{d}x^\nu}{\mathrm{d}\lambda}=f(r)\dfrac{\mathrm{d}t}{\mathrm{d}\lambda},\\[1em]
    L=&g_{\mu\nu}\xi_2^\mu\dfrac{\mathrm{d}x^\nu}{\mathrm{d}\lambda}=r^2\dfrac{\mathrm{d}\varphi}{\mathrm{d}\lambda}.\label{xi}
\end{split}
\end{equation}
Moreover, the photon’s 4‑momentum satisfies $g_{\mu\nu}K^\mu K^\nu=0$. Using the conserved quantities $E$ and $L$, the components of the photon 4‑momentum in the coordinates $\{t,r,\theta,\varphi\}$ are obtained as
\begin{equation}
    K^\mu=\left(\dfrac{E}{f(r)},\pm\sqrt{E^2-\dfrac{f(r)L^2}{r^2}},0,\dfrac{L}{r^2}\right),\label{gxk}
\end{equation}
where the plus sign corresponds to motion away from the BH and the minus sign to motion toward it. Simultaneously, using the radial component expression $K^r=\mathrm{d}r/\mathrm{d}\lambda$ presented in eq.~\eqref{gxk}, we derive the differential equation for the radial coordinate with respect to the affine parameter, which takes the form
\begin{equation}
    \left(\dfrac{\mathrm{d}r}{\mathrm{d}\lambda}\right)^2=E^2-f(r)\dfrac{L^2}{r^2}.\label{rl}
\end{equation}

\begin{figure*}[tp]  
        \centering  
            \includegraphics[width=0.45\textwidth]{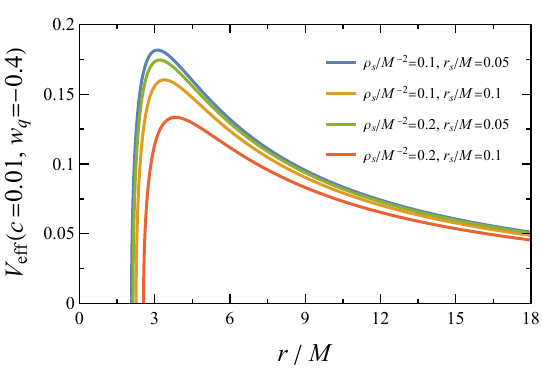}  
            \hspace{0.5cm}
            \includegraphics[width=0.45\textwidth]{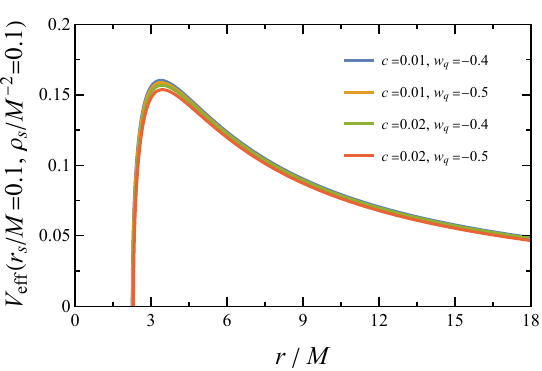}  
        \caption{Plots of the effective potential $V_\mathrm{eff}$ of photons versus the radial coordinate $r$ for different parameters. The left panel shows the variation with DM parameters ($\rho_s$, $r_s$) for fixed QF parameters ($c=0.01$, $w_q=-0.4$). The right panel shows the variation with QF parameters ($c$, $w_q$) for fixed DM parameters ($\rho_s/M^{-2}=0.1$, $r_s/M=0.1$).}  
        \label{veff}  
    \end{figure*}

Next, we discuss the effective potential for circular orbits of photons. For the circular orbits of photons, $\mathrm{d}r/\mathrm{d}\lambda=0$. Substituting this condition into eq.~\eqref{rl} and employing the definitions of the impact parameter and effective potential
\begin{equation}
    b:=\dfrac{L}{E},\qquad V_\mathrm{eff}:=\dfrac{1}{b},\label{bv}
\end{equation}
we obtain
\begin{equation}
    V_\mathrm{eff}(r)=\dfrac{\sqrt{f(r)}}{r}.\label{v}  
\end{equation}
Previous studies have separately investigated the properties of the effective potential in the presence of either a Dehnen-type DM halo~\citep{r78} or a QF~\citep{r85}. It was found that, under the exclusive influence of each component, increasing the respective parameters ($\rho_s$, $r_s$; $c$, $|w_q|$) suppresses the effective potential, shifts its extremum outward, and lowers the peak value.
To examine the influence of DM-DE interaction on the effective potential, we plot the photon effective potential for different parameter sets using eqs.~\eqref{v} and~\eqref{f}, as shown in figure~\ref{veff}. In all cases, the effective potential rises sharply from zero at the event horizon, reaches an extremum, and then gradually declines. The left panel reveals that, even in the presence of DE, the DM parameters $\rho_s$ and $r_s$ reduce the overall effective potential, shift the extremum away from the BH, and lower its value. Moreover, a larger $\rho_s$ (or $r_s$) enhances the sensitivity of the effective potential to changes in $r_s$ (or $\rho_s$). 
Similarly, the right panel shows that the QF parameters $c$ and $|w_q|$ also reduce the overall potential, shift the extremum outward, and decrease its value. However, their effect is weaker than that of DM, as shown in the left panel.
These conclusions can be reached through an analysis of eq.~\eqref{v}.
It should be emphasized that introducing the QF can alter the spacetime structure. Specifically, within the considered range ($-1<w_q<-1/3$), the spacetime described by eq.~\eqref{ds} changes from asymptotically flat to non-asymptotically flat, and features two horizons.  This leads to a steeper decline of the potential after the peak, as seen in figure~\ref{veff}, with it eventually falling to zero at the cosmological horizon (which lies outside the plotted range).

In figure~\ref{veff}, the extremum point $r_p$ of the effective potential for photons corresponds to a circular orbit of photons, satisfying
\begin{equation}
    \left.\dfrac{\partial V_\mathrm{eff}}{\partial r}\right|_{r=r_p}=0.
\end{equation}
Orbits satisfying this condition constitute the photon sphere. Photons on these orbits are, however, radially unstable. Consequently, any perturbation will inevitably deflect a photon, causing it to either spiral into the BH or scatter away to infinity.
Substituting the photon sphere radius $r_p$ into $b=1/V_\mathrm{eff}(r)$, The critical impact parameter $b_p$ is derived to be\newline
\begin{equation}
    b_p=\dfrac{r_p}{\sqrt{1-\dfrac{2M}{r_p}-32\pi\rho_sr_s^2\sqrt{\dfrac{r_p+r_s}{r_p}}-\dfrac{c}{r_p^{3w_q+1}}}}.
\end{equation}

To derive the trajectory of light rays, we start from the radial eq.~\eqref{rl}. By taking the square root and applying the chain rule $\mathrm{d}\lambda/\mathrm{d}\varphi$, we obtain
\begin{equation}  
    \dfrac{\mathrm{d}r}{\mathrm{d}\varphi}=\pm\dfrac{r^2}{b}\sqrt{1-f(r)\dfrac{b^2}{r^2}}.\label{rvarphi}  
\end{equation}  
The sign depends on the direction of light deflection and the variation of the radial coordinate. Introducing a new variable $u\equiv1/r$, eq.~\eqref{rvarphi} is transformed into 
\begin{equation}  
    \dfrac{\mathrm{d}u}{\mathrm{d}\varphi}\equiv\mp\sqrt{\mathnormal{\Omega}(u,b)}=\mp\sqrt{\dfrac{1}{b^2}-u^2f(u)},  
\end{equation}
where $\mathnormal{\Omega}(u,b)=1/b^2-u^2f(u)$. The $\mp$ in this equation corresponds to the $\pm$ in eq.~\eqref{rvarphi}. Separating variables and integrating gives  
\begin{equation}  
    \mp\int_0^\varphi\mathrm{d}\varphi=\int_{u_0}^u\dfrac{\mathrm{d}u}{\sqrt{\mathnormal{\Omega}(u,b)}},\label{xinvarphi}  
\end{equation}  
where the observer is assumed to be located at $(r_0=1/u_0=100M,\varphi_0=0)$. It should be emphasized that the behavior of $\mathnormal{\Omega}(u,b)$ depends crucially on the impact parameter $b$. Specifically, three distinct cases need to be addressed:
\begin{itemize}
    \item When $b = b_p$, we have $\mathnormal{\Omega}(1/r_p, b_p) = 0$. In this case, the integral is evaluated from $u_0$ to $1/r_p$ and diverges, indicating that light rays orbit the photon sphere indefinitely.
\end{itemize}
\begin{itemize}
    \item When $b < b_p$, the function $\mathnormal{\Omega}(u,b)$ possesses no roots in the interval $(u_0, 1/r_h)$, and the integration proceeds directly from $u_0$ to $1/r_h$.
\end{itemize}
\begin{itemize}
    \item When $b > b_p$, $\mathnormal{\Omega}(u,b)$ vanishes at some point $u_p$ lying between $u_0$ and $1/r_p$. Physically, this zero corresponds to the periastron of the light ray. In this case, $u$ increases from $u_0$ to $u_p$ (the turning point) and then decreases, requiring the integration domain to be split accordingly.
\end{itemize}
Thus, the choice of integration domain is determined by the value of $b$ relative to the critical impact parameter $b_p$.

\begin{figure*}[htp]
    \centering
        \includegraphics[width=0.32\textwidth]{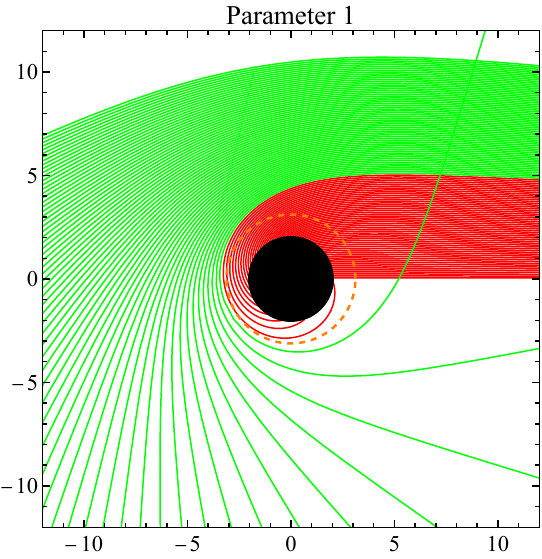}
        \includegraphics[width=0.32\textwidth]{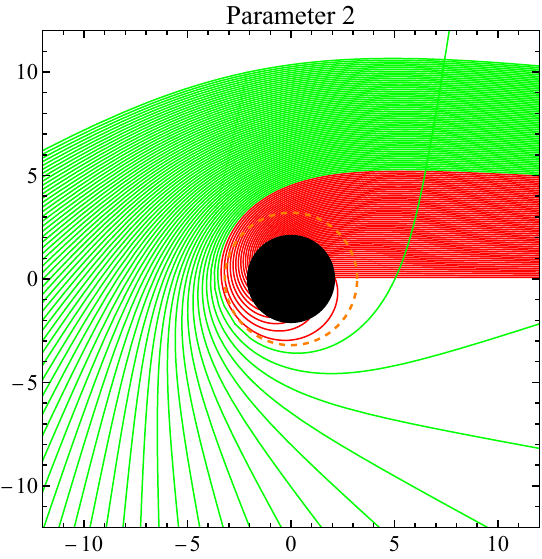}
        \includegraphics[width=0.32\textwidth]{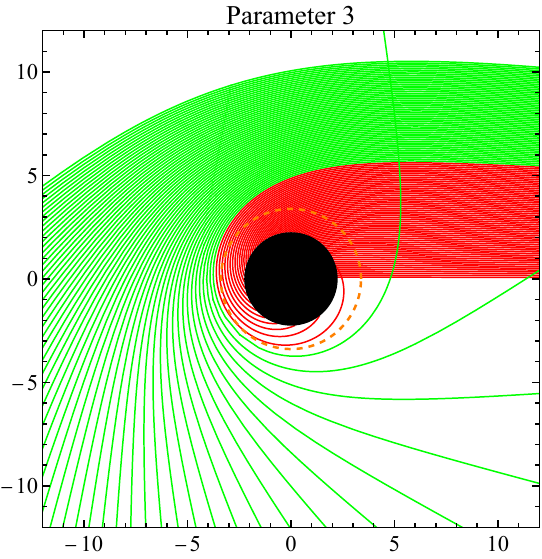}\\[1em]
        \includegraphics[width=0.32\textwidth]{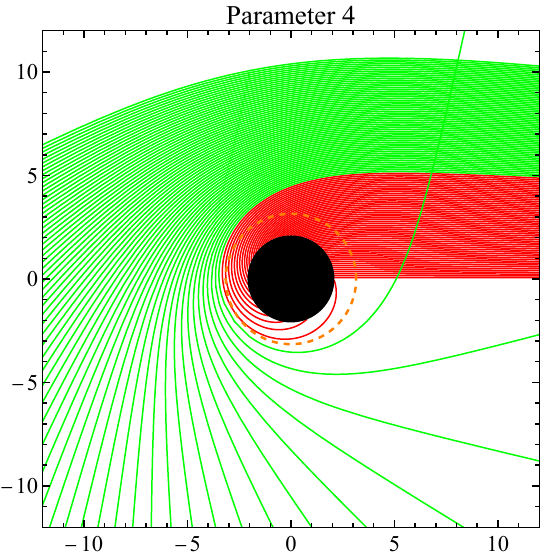}
        \includegraphics[width=0.32\textwidth]{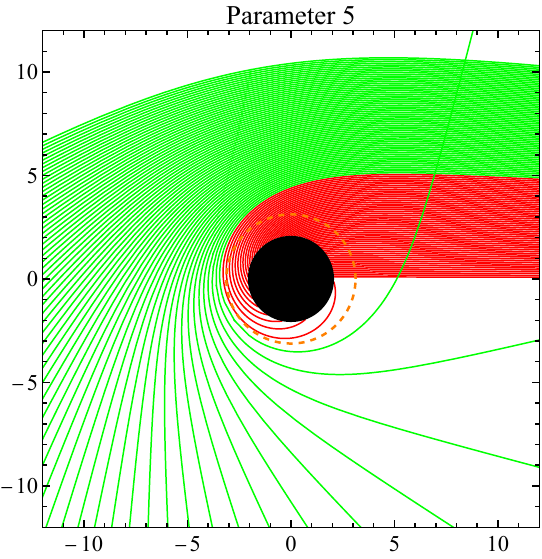}
    \caption{Light trajectories on the equatorial plane for BHs with different parameter sets (1–5 in table~\ref{canshu}).
    The horizontal and vertical axes of the figure represent the two-dimensional Cartesian coordinates $x$ and $y$ respectively, normalized by the BH mass $M$, where $x=(r/M)\cos\varphi$ and $y=(r/M)\sin\varphi$. The coordinate origin corresponds to the center of the BH, and $r$ and $\varphi$ are the second and fourth components of the static spherically symmetric coordinate system $(t, r, \theta, \varphi)$, respectively.
    Within each panel, light rays with $b<b_p$ and $b>b_p$ are depicted by red and green curves, respectively; the photon sphere is marked by an orange dashed line, and the event horizon appears as a black disk. All coordinates are scaled in units of the BH mass $M$.}
    \label{guiji}
\end{figure*}

Using a ray-tracing algorithm, we illustrate the light trajectories for BHs under parameter sets 1–5 (see tables~\ref{canshu} and \ref{jihecanshu} for the corresponding values. 
It is worth noting that these parameter sets are chosen to clearly demonstrate the effects of DM and DE on BH images within this theoretical analysis. In addition, the BH shadow radii $b_p/M$ computed in this work (see $b_p$ in the fourth row of table~\ref{jihecanshu}) all comply with the EHT observational constraint on the M87* shadow, $3.38\leq b_p/M\leq6.91$~\citep{d3}, verifying that the parameter selection is astrophysically reasonable.
) in figure~\ref{guiji}. Our calculations reveal that individually increasing the parameters, $\rho_s$, $r_s$, $c$, and $|w_q|$, within this composite background consistently enlarges the event horizon, photon sphere radius, and critical impact parameter. This enlargement relative to the Schwarzschild case ($r_h/M=2$, $r_p/M=3$, $b_p/M=3\sqrt{3}$) is consistent with established results for spacetimes containing a Dehnen-(1,4,5/2)-type DM halo or a QF in isolation~\citep{r78,r85}. Furthermore, the variations of the horizon and photon sphere are consistent with the potential analysis presented in figures~\ref{rhdm} and \ref{veff}.
\begin{table}
    \centering
    \begin{tabular}{cccccc}\hline
         Parameter&  1&  2&  3&  4& 5\\\hline
         $\rho_s/M^{-2}$&  $0.1$&  $0.2$&  $0.1$&  $0.1$& $0.1$\\
         $r_s/M$&  $0.05$&  $0.05$&  $0.1$&  $0.05$& $0.05$\\
         $c$&  $0.01$&  $0.01$&  $0.01$&  $0.02$& $0.01$\\
         $w_q$&  $-0.4$&  $-0.4$&  $-0.4$&  $-0.4$& $-0.5$\\ \hline
    \end{tabular}
    \caption{Selection of five different parameter sets. Parameter 1 is taken as the baseline set. Relative to this baseline, Parameter 2 has a higher DM halo density $\rho_s/M^{-2}$, Parameter 3 a larger DM halo scale radius $r_s/M$, Parameter 4 a larger value of the DE normalization factor $c$, and Parameter 5 a smaller (more negative) value of the DE equation-of-state $w_q$.}
    \label{canshu}
\end{table}
\begin{table}
    \centering
    \begin{tabular}{cccccc}\\\hline
         Parameter&  1&  2&  3&  4& 5\\\hline
         $r_h/M$&  2.077&  2.133&  2.259&  2.102& 2.083\\
         $r_p/M$&  3.114&  3.199&  3.387&  3.152& 3.121\\
         $b_p/M$&  5.506&  5.731&  6.239&  5.616& 5.550\\ \hline
    \end{tabular}
    \caption{Values of the event horizon, photon sphere, and critical impact parameter for each parameter set.}
    \label{jihecanshu}
\end{table}

\section{Shadows and photon rings of BHs embedded in a Dehnen-(1,4,5/2)-type DM halo with a QF under different spherical accretion models\label{sub4}}
Matter in the universe, such as plasma, gas, and dust, often exists in a diffuse state. When such matter approaches a BH, it gradually accumulates under gravitational influence to form an accretion flow. During accretion, the motion of matter releases radiation, which provides key evidence for observers to detect optical phenomena such as BH shadows and photon rings. By studying the characteristics of BH shadows and photon rings under different types of spherical accretion, we can investigate how accretion processes affect BH images, and further explore the properties of Dehnen-(1,4,5/2) DM and QF surrounding BHs. In this section, assuming the presence of abundant radiating matter in the universe, we analyze how the size and intensity of BH shadows and the morphology of photon rings are influenced by Dehnen-(1,4,5/2) DM, QF, and the kinematic properties of accreting matter under both static and infalling accretion scenarios.

In spherical accretion, the specific intensity at the observer's frequency $\nu_o$, denoted as $I_\mathrm{obs}(b,\nu_o)$, can be expressed as~\citep{r48,r94,r95,r96}
\begin{equation}
    I_\mathrm{obs}(b,\nu_o)=\int_\mathrm{ray}g^3j(\nu_e)\mathrm{d}l.\label{iobs}
\end{equation}
Here, $g=\nu_o/\nu_e$ is the redshift factor, representing the ratio of the photon frequency measured by the observer at the observation point $\nu_o$ to the photon frequency measured by an observer comoving with the accreting material $\nu_e$. $j(\nu_e)$ is the volume emissivity of the accreting material. We adopt the simplified monochromatic radiation scenario suggested in reference~\citep{r48}, where the radial distribution of emissivity follows a $1/r^2$ scaling and the frequency is fixed at $\nu$, i.e., $j(\nu_e)\propto\delta(\nu_e-\nu)/r^2$. $\mathrm{d}l$ is the proper distance along the null geodesic in the comoving frame of the accretion flow. Finally, the integration path “ray” denotes the complete trajectory of the light rays. The integrated intensity $F_\mathrm{obs}$ is obtained by integrating the specific intensity over all frequencies
\begin{equation}
    F_\mathrm{obs}(b)=\int_0^{+\infty}I_\mathrm{obs}(b,\nu_o)\mathrm{d}\nu_o.\label{fobs}
\end{equation}

\begin{figure*}[htp]  
    \centering  
    \includegraphics[width=0.48\textwidth]{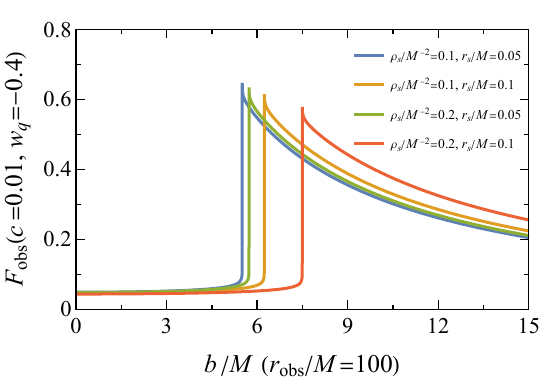} 
    \hspace{0.2cm}
    \includegraphics[width=0.48\textwidth]{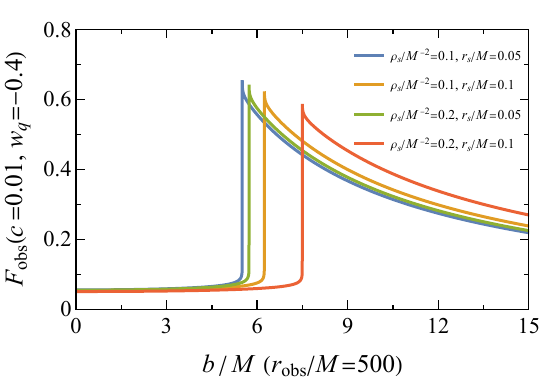}  
    \includegraphics[width=0.48\textwidth]{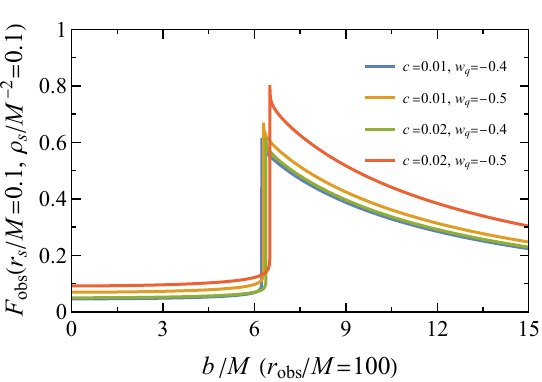}
    \hspace{0.2cm}
    \includegraphics[width=0.48\textwidth]{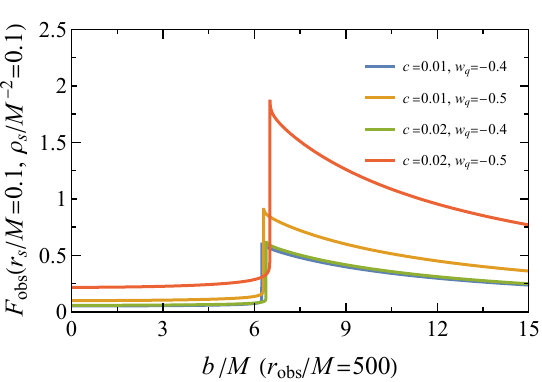}  
    \includegraphics[width=0.48\textwidth]{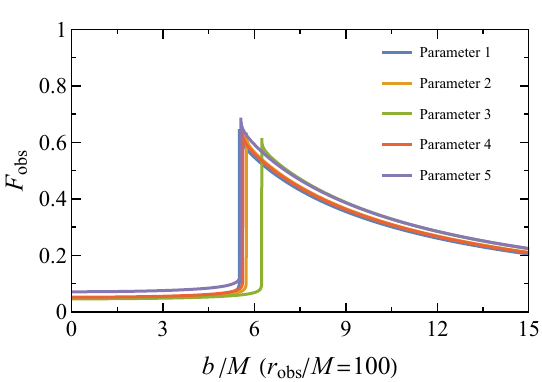}
    \hspace{0.1cm}
    \includegraphics[width=0.49\textwidth]{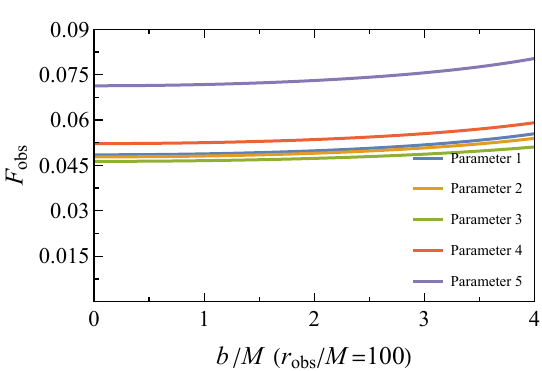}  
    \caption{Dependence of the observed integrated intensity $F_\mathrm{obs}$ on the impact parameter $b$ for different spacetime parameters and observer positions, under infalling spherical accretion.}
    \label{fobsxx}  
\end{figure*}

To solve for the redshift factor, we need to analyze the four-velocities of the observer and the accretion flow.  
Denote the four-velocity of the accreting matter as $Z^\mu$, and the four-velocity of a static observer as $U^\mu$.  
Since the four-velocity of a static observer satisfies $g_{\mu\nu}U^\mu U^\nu=-1$, we have  
\begin{equation}
    U^\mu=\left(\dfrac{1}{\sqrt{f(r_\mathrm{obs})}},0,0,0\right),\label{uobs}
\end{equation}
where $r_\mathrm{obs}$ is the radial coordinate of the observer.  
The four-velocity $Z^\mu$ of the accreting matter also satisfies $g_{\mu\nu}Z^\mu Z^\nu=~-1$. 
In the static accretion, similarly,  
\begin{equation}
    Z^\mu=\left(\dfrac{1}{\sqrt{f(r)}},0,0,0\right).\label{jz}
\end{equation}
For infalling accretion, starting from the normalization condition of the timelike 4-velocity, together with the fact that the 4-velocity $Z^\mu$ of the infalling accretion flow has only non-vanishing temporal and radial components, as well as the conserved specific energy $\mathscr{E}$ along the fluid worldline, we obtain
\begin{equation}
    Z^\mu=\left(\dfrac{\mathscr{E}}{f(r)},-\sqrt{\mathscr{E}^2-f(r)},0,0\right),\label{xz}
\end{equation}
where the minus sign in the radial component is adopted to account for the inward motion of the accreting matter toward the BH.
The quantity $\mathscr{E}$ in eq.~\eqref{xz} is the conserved energy along the geodesic of the accretion flow, given by $\mathscr{E}=-g_{\mu\nu}Z^\mu\xi_1^\nu$, and can be determined by the initial position from which the matter starts infalling from rest.  
If the accretion matter begins to fall from rest at $r=r_\mathrm{ini}$, where its four-velocity is $(f^{1/2}(r_\mathrm{ini}),0,0,0)$, we obtain $\mathscr{E}=f^{1/2}(r_\mathrm{ini})$.  
It is worth noting that in spacetimes with a cosmological horizon, the initial position of the accretion matter and the observer must lie between the BH event horizon and the cosmological horizon. 
We emphasize that if $\mathscr{E}=f^{1/2}(r_\mathrm{ini})=f^{1/2}(r)$, eq.~\eqref{xz} reduces to eq.~\eqref{jz} at the location of the accreting matter, i.e., it returns to the static accretion case.  
Thus, static accretion is a limiting case of infalling accretion, corresponding to the scenario where the radial inflow velocity of the accretion flow is identically zero at all times and spatial positions.
The redshift factor can be expressed by the 4-velocities of the observer and the accreting matter
\begin{equation}
    g:=\dfrac{\nu_o}{\nu_e}=\dfrac{(-K^\mu U_\mu)|_{r_\mathrm{obs}}}{(-K^\nu Z_\nu)|_{r_\mathrm{em}}},\label{g}
\end{equation}
where the subscript ``$r_\mathrm{em}$'' denotes the photon emission point of the accreting matter. Finally, the proper distance $\mathrm{d}l$ is defined as~\citep{r51,r94,r95}
\begin{equation}
    \mathrm{d}l:=\nu_e\mathrm{d}\lambda=-K_\mu Z^\mu\mathrm{d}\lambda.\label{dl}
\end{equation}

\subsection{Shadow and photon ring under infalling spherical accretion}
Using eqs.~\eqref{gxk},~\eqref{uobs},~\eqref{xz} and~\eqref{g}, the redshift factor for infalling accretion is obtained as
\begin{equation}
    g_\pm=\dfrac{f(r)}{\sqrt{f(r_\mathrm{obs})}}\dfrac{1}{\sqrt{f(r_\mathrm{ini})}\pm\sqrt{(1-b^2f(r)/r^2)(f(r_\mathrm{ini})-f(r))}},\label{xh}
\end{equation}
where the $\pm$ corresponds to the $\pm$ in eq.~\eqref{gxk}. Then, with eqs.~\eqref{iobs},~\eqref{fobs} and~\eqref{dl}, the expression for the integrated intensity of infalling accretion is
\begin{equation}
    F_\mathrm{obs}(b)=\int\dfrac{g_\pm^4}{r^2f(r)}\left(\sqrt{f(r_\mathrm{ini})-f(r)}\pm r\sqrt{\dfrac{f(r_\mathrm{ini})}{r^2-b^2f(r)}}\right)\mathrm{d}r.\label{fobsx}
\end{equation}
For convenience, the infall initial position $r_\mathrm{ini}$ of the accretion flow is chosen as
\begin{equation}
    \left.\dfrac{\partial f(r)}{\partial r}\right|_{r=r_{\mathrm{ini}}}=0,
\end{equation}
which ensures the radial coordinate component in eq.~\eqref{xz} takes meaningful values. Figure~\ref{fobsxx} shows the functional relation between the integrated intensity $F_\mathrm{obs}$ and the impact parameter $b$ for infalling spherical accretion in spacetimes with different parameters and for observers at different locations. It can be seen that all integrated intensity curves exhibit a region of very low intensity—the shadow region—followed by a rapid increase as $b$ grows and then a slow decline. In such cases, the photon ring radius equals the critical impact parameter. The first row illustrates the influence of different observer positions and different DM parameters $\rho_s$ and $r_s$ on the integrated intensity under the QF. The left panel corresponds to an observer at $r_\mathrm{obs}/M=100$, the right panel to an observer at $r_\mathrm{obs}/M=500$. The impact of $\rho_s$ and $r_s$ on the integrated intensity is nearly independent of the observer’s location. Furthermore, larger $\rho_s$ (or $r_s$) makes the integrated intensity more sensitive to variations in $r_s$ (or $\rho_s$), and both parameters enlarge the shadow region. The integrated intensity within the shadow is almost identical for different parameter values.

\begin{figure*}[htp]
    \centering
        \includegraphics[width=0.32\textwidth]{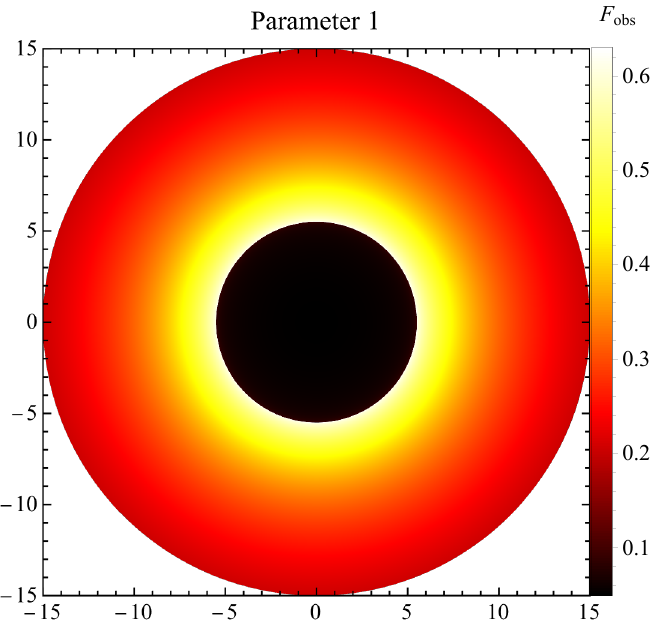}
        \includegraphics[width=0.32\textwidth]{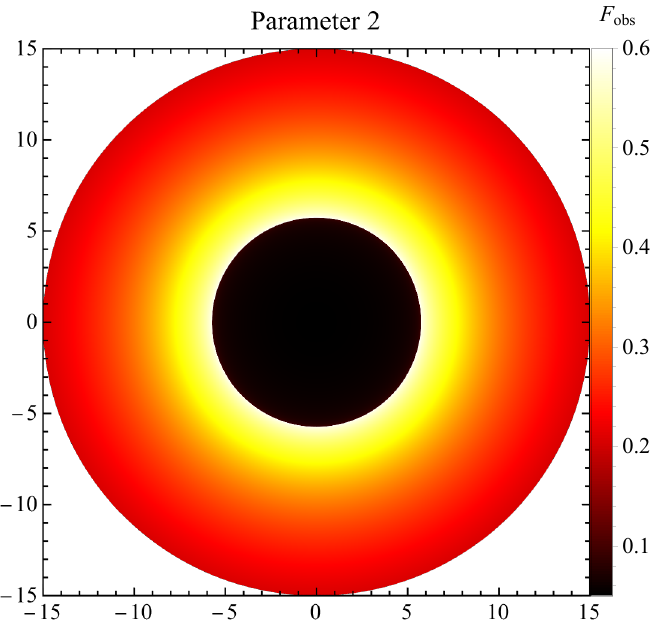}
        \includegraphics[width=0.32\textwidth]{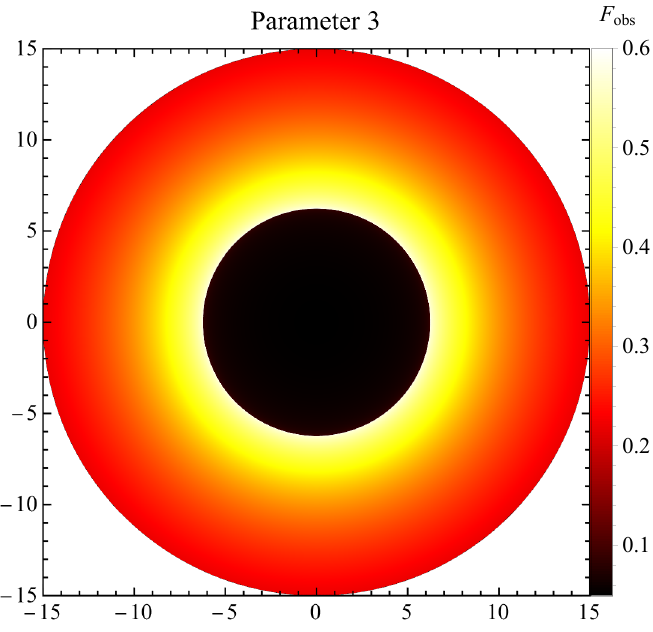}\\[1em]
        \includegraphics[width=0.32\textwidth]{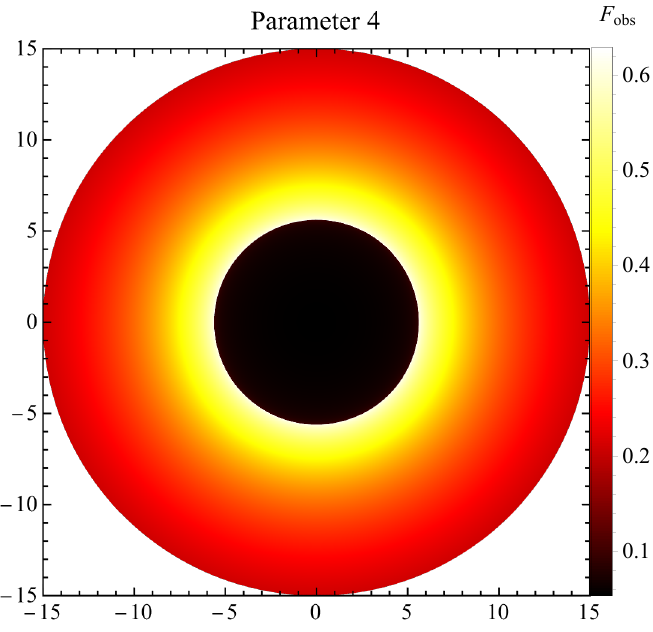}
        \includegraphics[width=0.32\textwidth]{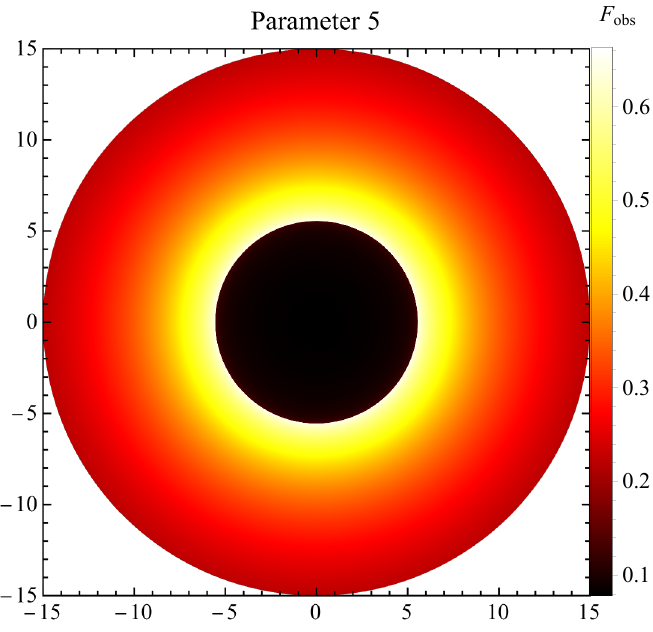}
    \caption{BH images under a radially infalling spherical accretion flow, corresponding to the parameter sets shown in the third row of figure~\ref{fobsxx}.}
    \label{bhx}
\end{figure*}

The second row in figure~\ref{fobsxx} presents the influence of different observer positions and different QF parameters $c$ and $w_q$ on the integrated intensity under the Dehnen-(1,4,5/2) DM profile. Evidently, the impact of $c$ and $w_q$ on the integrated intensity strongly depends on the observer’s position. Observers at larger distances generally record higher observed intensities. Compared with the subplots in the first row, a larger $c$ (or $w_q$) makes the intensity in the shadow region more sensitive to variations in $w_q$ (or $c$), and generally enhances the shadow intensity (integrated intensity in the shadow), while having little effect on the size of the shadow region.  
The third row of figure~\ref{fobsxx} shows the integrated intensity images for the five different parameter sets listed in table~\ref{canshu}. Figure~\ref{bhx} presents the BH images corresponding to these parameters (Parameter sets 1–5). 
As can be seen from the right panel in the third row of figure~\ref{fobsxx}, larger values of the DM halo parameters $\rho_s$ and $r_s$ lead to a decrease in the shadow intensity, while larger values of the QF parameters $c$ and $w_q$ result in an increase in the shadow intensity.

\subsection{Shadow and photon ring under static spherical accretion}
As previously stated, static accretion is the zero velocity limiting case of infalling accretion. Accordingly, by taking the limit where $r\to r_\mathrm{ini}$ in eq.~\eqref{xh}, we obtain the redshift factor for static accretion as
\begin{equation}
    g=\sqrt{\dfrac{f(r)}{f(r_\mathrm{obs})}}.\label{jhy}
\end{equation}
Then, by taking the limit where $r\to r_\mathrm{ini}$ in eq.~\eqref{fobsx} and substituting eq.~\eqref{jhy} into the resulting expression, we obtain the integrated intensity in the static accretion model
\begin{equation}
    F_\mathrm{obs}(b)=\int\dfrac{f(r)}{rf^2(r_\mathrm{obs})}\sqrt{\dfrac{f(r)}{r^2-b^2f(r)}}\mathrm{d}r.\label{fobsjj}
\end{equation}

\begin{figure*}[htp]
    \centering
        \includegraphics[width=0.48\textwidth]{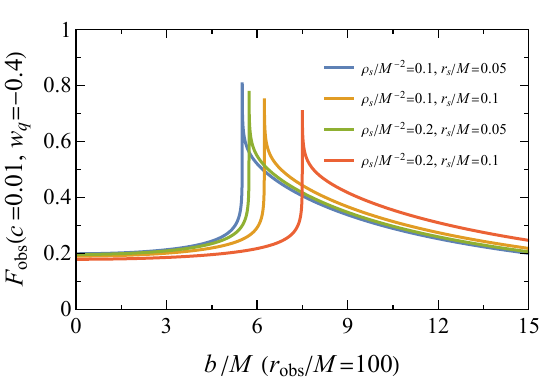}
        \hspace{0.2cm}
        \includegraphics[width=0.48\textwidth]{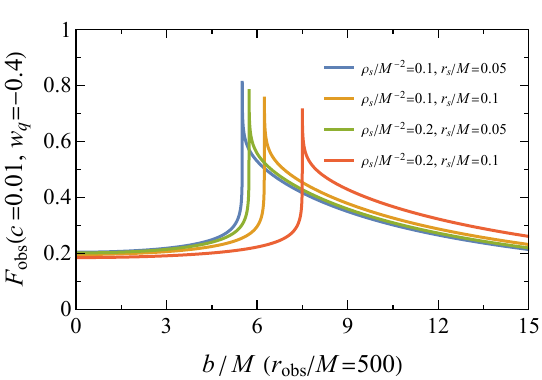}
        \includegraphics[width=0.48\textwidth]{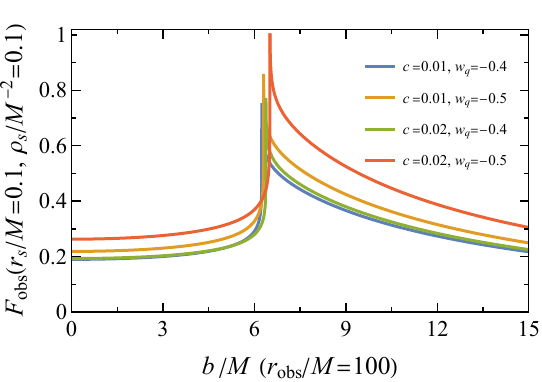}
        \hspace{0.2cm}
        \includegraphics[width=0.48\textwidth]{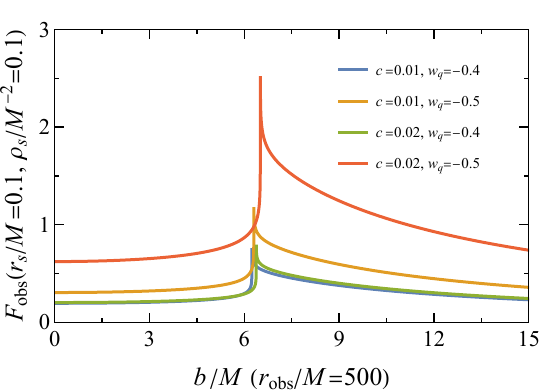}
        \includegraphics[width=0.48\textwidth]{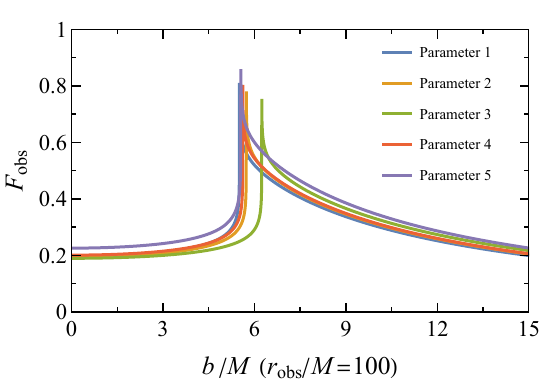}
        \includegraphics[width=0.49\textwidth]{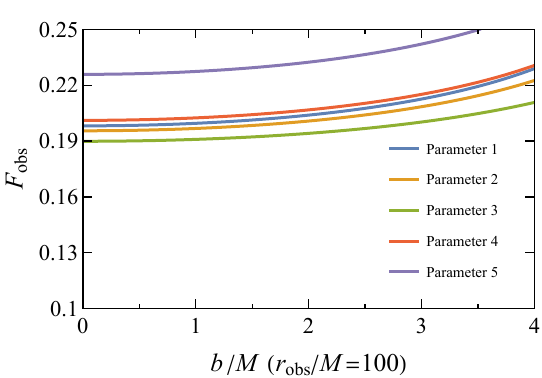}
    \caption{Dependence of the observed integrated intensity $F_\mathrm{obs}$ on the impact parameter $b$ for different spacetime parameters and observer positions, under static spherical accretion.}
        \label{fobsj}
\end{figure*}

Figure~\ref{fobsj} shows the relationship between the integrated intensity $F_\mathrm{obs}$ and the impact parameter $b$ for a BH under static spherical accretion, for various parameters and observer positions. Similar to the case of infalling accretion, regardless of the chosen parameters, the variation of integrated intensity follows the same pattern: as $b$ increases from zero, $F_\mathrm{obs}$ rises rapidly from a small value, peaks at the critical impact parameter $b_p$ corresponding to the given parameters, and then decreases slowly. In this static scenario, the photon ring radius again coincides with $b$. This confirms a key result: under spherical accretion, the photon ring radius is strictly equal to the critical impact parameter. Therefore, it is determined solely by the background spacetime geometry and is independent of the accretion flow’s kinematic details (e.g., static and infalling).

The first row of subplots in figure~\ref{fobsj} shows the influence of different observer positions and different DM parameters $\rho_s$ and $r_s$ on the integrated intensity under a QF. Similar to the infall accretion, the effect of $\rho_s$ and $r_s$ on the integrated intensity is almost independent of the observer's position. A larger $\rho_s$ (or $r_s$) leads to a stronger sensitivity of the integrated intensity to $r_s$ (or $\rho_s$), and both cause an enlargement of the shadow region. The peak value of the integrated intensity in the shadow boundary is nearly identical across different parameters, though the variation in the profile shape is more pronounced in this static scenario than in the infall accretion case.

The second row in figure~\ref{fobsj} shows the influence of different observer positions and different QF parameters $c$ and $w_q$ on the integrated intensity under Dehnen-(1,4,5/2) DM. Similar to the infalling accretion, the influence of QF parameters on the integrated intensity strongly depends on the observer position. More distant observers generally have higher integrated intensity. A larger value of $c$ (or $w_q$) makes the shadow intensity more sensitive to $w_q$ (or $c$ respectively), enhances the shadow intensity, but has a negligible effect on the BH shadow size.

\begin{figure*}[htp]  
    \centering  
        \includegraphics[width=0.32\textwidth]{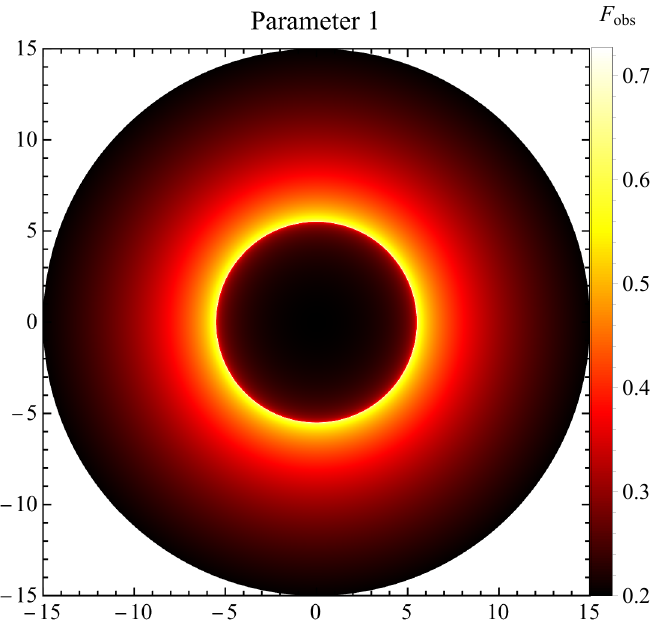}
        \includegraphics[width=0.32\textwidth]{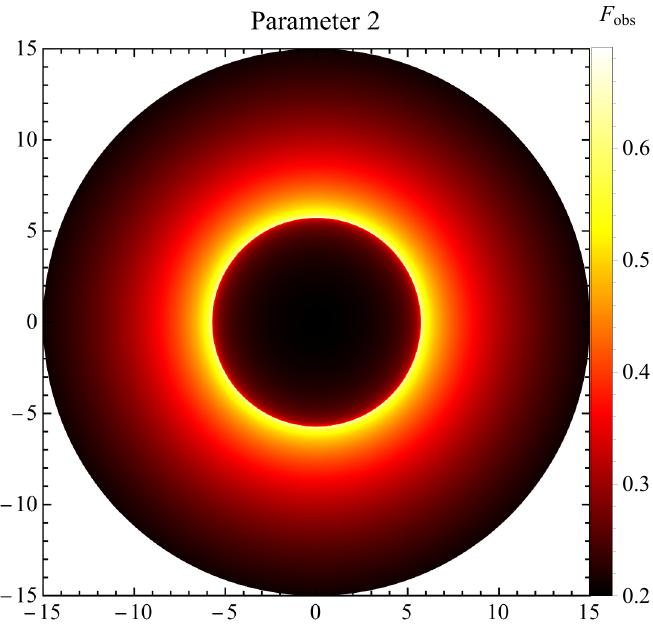}  
        \includegraphics[width=0.32\textwidth]{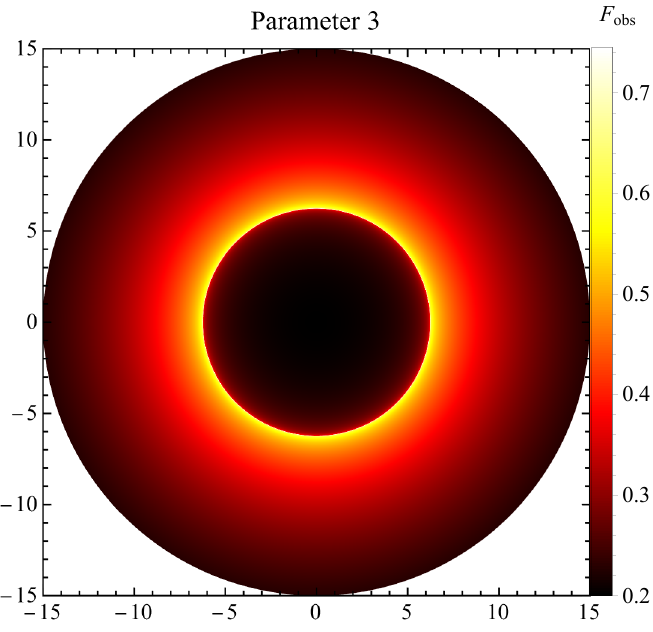}\\[1em]  
        \includegraphics[width=0.32\textwidth]{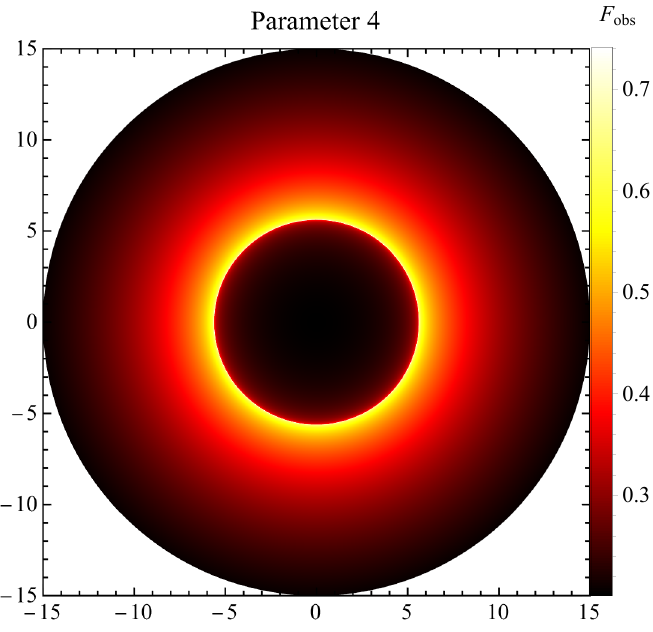}  
        \includegraphics[width=0.32\textwidth]{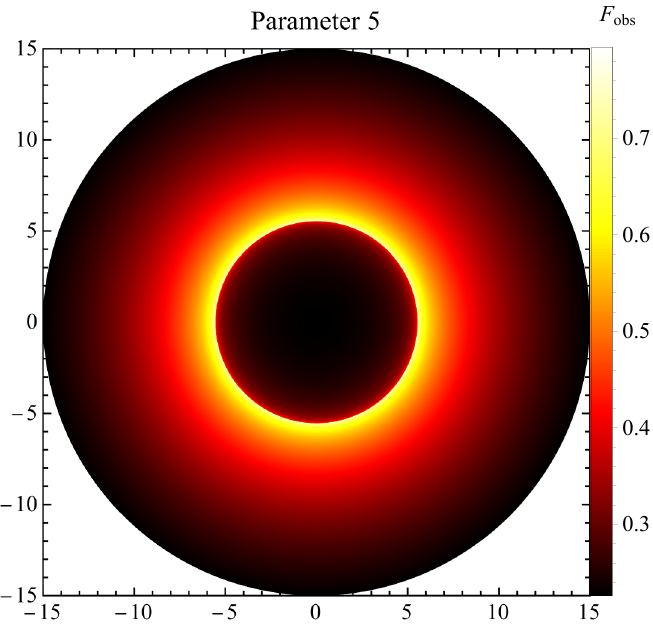}  
    \caption{BH images under a radially static spherical accretion flow, corresponding to the parameter sets shown in the third row of figure~\ref{fobsj}. The selected parameters are the same as in figure~\ref{bhx}.}  
    \label{bhj}  
\end{figure*}

Figure~\ref{fobsj} shows in its third row the integrated intensity maps for five different parameter sets under static accretion, corresponding to the parameters in table~\ref{canshu}. Figure~\ref{bhj} displays the BH images for the respective parameters (Parameter sets 1–5).
As can be seen from the right panel in the third row of figure~\ref{fobsj}, similar to infalling accretion, larger DM parameters $\rho_s$ and $r_s$ lead to a decrease in shadow intensity, while larger QF parameters $c$ and $w_q$ cause an increase in shadow intensity.
This distinct response offers a potential diagnostic: based on whether the shadow appears dimmer or brighter, one could infer whether DM or DE dominates the interaction with the BH. Finally, it is worth noting that the integrated intensity in the shadow in static accretion is much higher than that in infalling accretion. The reason for this is that the static accretion flow has no radial velocity relative to the BH, so there is no Doppler redshift to dim the emission.

\subsection{Physical mechanism for the distinct observer-distance dependences of DM and DE effects}

In this section, we provide a physical explanation for the pronounced difference in observer-distance dependence between DM and DE effects.

From the metric, we can derive the energy-momentum tensor $T_{\mu\nu}$ of this spacetime via the Einstein field equations $G_{\mu\nu}=8\pi T_{\mu\nu}$, where the Einstein tensor is defined as
\begin{equation}
    G_{\mu \nu} \equiv R_{\mu\nu}-\dfrac{1}{2}R g_{\mu\nu}.
\end{equation}
Here $R_{\mu\nu}$ is the Ricci tensor, defined as $R_{\mu\nu}\equiv R_{\mu\sigma\nu}{}^{\sigma}$; $R$ is the Ricci scalar, defined as $R\equiv g^{\mu\nu}R_{\mu\nu}$; and $R_{\mu\nu\sigma}{}^{\rho}$ is the Riemann tensor, defined as
\begin{equation}
    {R_{\mu\nu\sigma}}^{\rho}\equiv\mathnormal{\Gamma}^{\rho }{}_{\sigma \mu,\nu }-\mathnormal{\Gamma}^{\rho}{}_{\sigma\nu,\mu}+\mathnormal{\Gamma}^{\lambda}{}_{\sigma\mu}\mathnormal{\Gamma}^{\rho}{}_{\nu\lambda}-\mathnormal{\Gamma}^{\lambda}{}_{\sigma\nu}\mathnormal{\Gamma}^{\rho}{}_{\mu\lambda}.
\end{equation}
In the above expression, $\mathnormal{\Gamma}^{\sigma}{}_{\mu \nu}$ is the Christoffel symbol, defined as
\begin{equation}
    \mathnormal{\Gamma}^{\sigma}{}_{\mu\nu} \equiv \frac{1}{2} g^{\sigma \lambda}\left(g_{\lambda \mu, \nu}+g_{\nu \lambda, \mu}-g_{\mu \nu, \lambda}\right).
\end{equation}
Substituting the metric into the above relations yields the components of the Ricci tensor:
\begin{equation}
    \begin{aligned}
    R_{tt}&=\dfrac{1}{2}f(r)f''(r)+\dfrac{f(r)f'(r)}{r},\\[0.3em]
    R_{rr}&=-\dfrac{f''(r)}{2f(r)}-\dfrac{f'(r)}{rf(r)},\\[0.3em]
    R_{\theta\theta}&=1-f(r)-rf'(r), \\[0.3em]
    R_{\varphi\varphi}&=\left(1-f(r)-rf'(r)\right)\sin^2\theta,
    \end{aligned}
\end{equation}
where $f'(r)=\mathrm{d}f(r)/\mathrm{d}r$ and $f''(r)=\mathrm{d}^2f(r)/\mathrm{d}r^2$. The corresponding Ricci scalar is then given by
\begin{equation}
    R=\dfrac{2}{r^2}-f''(r)-\dfrac{4f'(r)}{r}-\dfrac{2f(r)}{r^2}.
\end{equation}
Substituting into the Einstein field equations gives the energy-momentum tensor components
\begin{equation}
    \begin{aligned}
    T_{tt}&=\dfrac{G_{tt}}{8\pi}=\dfrac{1}{8\pi}\left(\dfrac{f(r)}{r^2}-\dfrac{f(r)f'(r)}{r}-\dfrac{f^2(r)}{r^2}\right), \\
    T_{rr}&=\dfrac{G_{rr}}{8\pi}=\dfrac{1}{8\pi}\left(\dfrac{f'(r)}{rf(r)}+\dfrac{1}{r^2}-\dfrac{1}{r^2f(r)}\right), \\
    T_{\theta\theta}&=\dfrac{G_{\theta\theta}}{8\pi}=\dfrac{1}{8\pi} \left(rf'(r)+\dfrac{r^2f''(r)}{2}\right),\\
    T_{\varphi\varphi}&=\dfrac{G_{\varphi\varphi}}{8\pi}=\dfrac{1}{8\pi} \left[\left(rf'(r)+\dfrac{r^2f''(r)}{2}\right)\sin^2\theta\right].
\end{aligned}
\end{equation}

To obtain the cosmic density and pressure in accordance with the standard perfect-fluid cosmological framework, we adopt an orthonormal tetrad basis $\{(e_{\mu})^a\}$
\begin{equation}
\begin{aligned}
    e_t^{\mu}=\dfrac{1}{\sqrt{f(r)}}\updelta_t^{\mu},\qquad
    e_r^{\mu}=\sqrt{f(r)}\updelta_r^{\mu},\\
    e_{\theta}^{\mu}=\dfrac{1}{r}\updelta_{\theta}^{\mu},\qquad
    e_{\varphi}^{\mu}=\dfrac{1}{r\sin\theta}\updelta_{\varphi}^{\mu}.
\end{aligned}
\end{equation}
From this we obtain the physical fluid quantities
\begin{equation}
\begin{aligned}
    \rho=T_{\mu\nu}e_t^{\mu}e_t^{\nu},\qquad
    p_r=T_{\mu\nu}e_r^{\mu}e_r^{\nu},\\[0.1em]
    p_{\theta}=T_{\mu\nu}e_{\theta}^{\mu}e_{\theta}^{\nu},\qquad
    p_{\varphi}=T_{\mu\nu}e_{\varphi}^{\mu}e_{\varphi}^{\nu},
\end{aligned}
\end{equation}
which reduce to
\begin{equation}
    \begin{aligned}
        \rho=&-p_r=\dfrac{1}{8\pi}\left(\dfrac{1}{r^2}-\dfrac{f'(r)}{r}-\dfrac{f(r)}{r^2}\right),\\
        p_{\theta}=&p_{\varphi}=\dfrac{1}{8\pi}\left(\dfrac{f'(r)}{r}+\dfrac{f''(r)}{2} \right).
    \end{aligned}
\end{equation}
According to the Friedmann acceleration equation, on cosmic scales the acceleration of the scale factor $a(t)>0$ satisfies
\begin{equation}
    3\ddot{a}=-4\pi a(\rho+p_r+p_\theta+p_\varphi)=-8\pi ap_{\theta},
\end{equation}
indicating that the evolution of $a(t)$ is governed by $p_{\theta}$ (or equivalently $p_{\varphi}$). From the explicit form of the metric function~\eqref{f}, we have
\begin{equation}
    p_{\theta}(r)=\dfrac{\rho_s}{2}\left(\dfrac{r_s}{r}\right)^4\left(\dfrac{r}{r+r_s}\right)^{3/2}-\dfrac{3cw_q\left(1+3w_q\right)}{16\pi r^{3(w_q+1)}}.
\end{equation}

When only the Dehnen-(1,4,5/2) DM or only QF is considered, the transverse pressure takes the form
\begin{equation}
    p_{\theta}(r)=\begin{cases}
    \displaystyle\dfrac{\rho_s}{2}\left(\dfrac{r_s}{r}\right)^4\left(\dfrac{r}{r+r_s}\right)^{3/2},&c=0,\\[1em]
    \displaystyle
    -\dfrac{3cw_q(1+3w_q)}{16\pi r^{3(w_q+1)}},&\rho_s=0.
\end{cases}
\end{equation}
It can be seen from the above that
\begin{equation}
    \begin{aligned}
        p_{\theta}&\geq0,\qquad\rho_s>0,\,r_s>0,\,c= 0,\\
        p_{\theta}&< 0,\qquad\rho_s = 0,\,c > 0,\,w_q < -\dfrac{1}{3}.
    \end{aligned}
\end{equation}
Thus, in the absence of QF, $p_{\theta}$ remains positive everywhere, leading to $\ddot{a}(t)<0$. The universe does not undergo accelerated expansion, and the causal structure of spacetime remains unchanged, so no cosmological horizon is formed. Furthermore, $p_{\theta}$ decreases rapidly with increasing radial coordinate, because the Dehnen-(1,4,5/2) DM halo is a localized matter distribution whose density falls off sharply with radius. It modifies the spacetime geometry only in the vicinity of the BH, within a few scale radii $r_s$. For an observer located far outside the halo ($r_{\text{obs}}\gg r_s$), spacetime returns to a nearly flat state. After photons propagating from the BH to distant observers exit the halo region, the additional gravitational effect from the DM halo becomes negligible. As a result, the observed shadow intensity are almost independent of the observer’s radial position.

In contrast, when the Dehnen-(1,4,5/2) DM is neglected, $p_{\theta}$ remains negative for $w_q<-1/3$, giving $\ddot{a}(t)>0$ and driving accelerated cosmic expansion, with sufficiently strong negative pressure for the parameters considered here. In fact, QF is a cosmic-scale energy component. Moreover, according to the above expression, $-p_{\theta}$ increases with radial coordinate $r$ when $w_q<-1$. The special property that $p_\theta<0$ for $w_q<-1/3$ in QF fundamentally alters the asymptotic structure of spacetime and introduces a cosmological horizon, so that spacetime is no longer asymptotically flat.

The introduction of a cosmological horizon and the breakdown of asymptotic flatness completely modify the behavior of the metric function: specifically, the metric function changes from monotonically increasing to first rising and then falling, eventually dropping to zero at the cosmological event horizon. This implies that photons propagating from the BH to the observer continuously accumulate gravitational redshift induced by the QF along the entire path. As can be seen from the redshift factor expression~\eqref{xh}, due to this structural change of the metric function, a more distant observer corresponds to a larger redshift factor and a stronger cumulative redshift effect, resulting in a higher observed intensity — the integrated intensity in equation~\eqref{fobsx} is proportional to the fourth power of the redshift factor. This is precisely the physical origin of the observer-distance dependence of the DE effect.

We will encounter the same phenomenon in section~\ref{sub5}, whose underlying physical mechanism is consistent with the analysis presented in this subsection.

\section{Shadows and photon rings of BHs embedded in a Dehnen-(1,4,5/2)-type DM halo with a QF under different thin-disk models\label{sub5}}
This section explores the pervasive phenomenon of accretion disks in the universe. In this accretion model, disk-like accretion flows serve as the light source of BHs. A static, optically thin, and geometrically thin accretion disk is assumed to exist on the equatorial plane, with the observer located directly above the disk plane. We then separately analyze light rays classification, transfer functions, and three specific thin‑disk models to reveal the image characteristics of supermassive BHs embedded in a Dehnen-(1,4,5/2) DM halo with a QF.

\subsection{Direct emission, lensed ring emission, and photon ring emission}
Light rays emitted from the accretion disk are deflected by the strong gravity of the BH, ultimately being captured by the BH, escaping to infinity (asymptotically flat spacetime), or entering the cosmological horizon (in non‑asymptotically flat spacetime with a cosmological horizon).
Before reaching the observer, a light ray may cross the accretion disk multiple times. 
Using eq.~\eqref{xinvarphi}, the azimuthal angle variation of a ray with $k$ intersections is obtained as $\Delta\varphi=\Delta\varphi_k=\pi/2+(k-1)\pi$.
The total photon orbit number is
\begin{equation}
    \Delta n=\dfrac{\Delta\varphi}{2\pi}.
\end{equation}
Inspired by the classification framework pioneered by Gralla et al.~\citep{r54}, we refine their criteria to categorize photons from a thin accretion disk into three observable types based on the orbit number $\Delta n$, with the classification of direct emission specifically adjusted. Concretely,
\begin{itemize}
    \item When $0.25\leq\Delta n<0.75$, the ray intersects the accretion disk only once, corresponding to direct emission;
\end{itemize}
\begin{itemize}
    \item When $0.75\leq\Delta n<1.25$, the ray intersects the accretion disk twice, corresponding to lensed ring emission;
\end{itemize}
\begin{itemize}
    \item When $\Delta n\geq1.25$, the ray intersects the accretion disk three or more times, corresponding to photon ring emission.
\end{itemize}

\begin{figure*}[htp]
    \centering
        \includegraphics[width=0.48\textwidth]{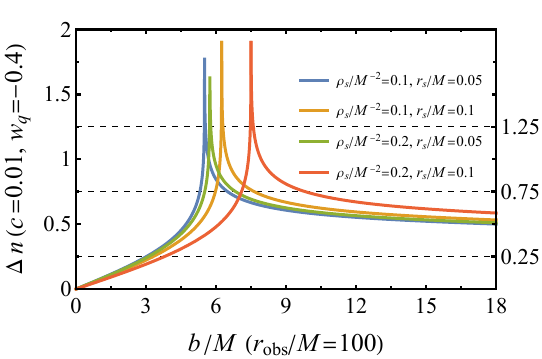}
        \hspace{0.2cm}
        \includegraphics[width=0.48\textwidth]{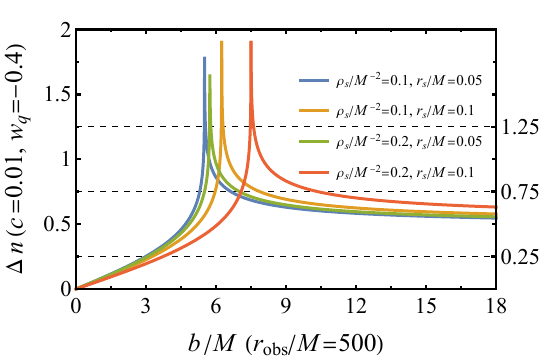}
        \includegraphics[width=0.48\textwidth]{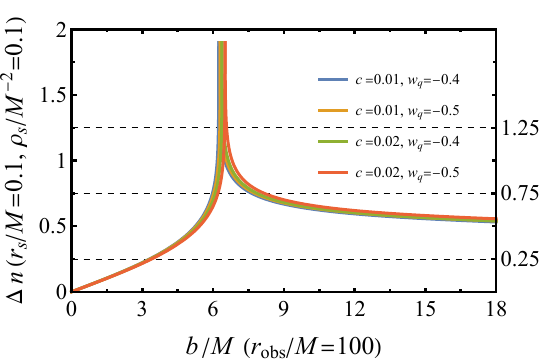}
        \hspace{0.2cm}
        \includegraphics[width=0.48\textwidth]{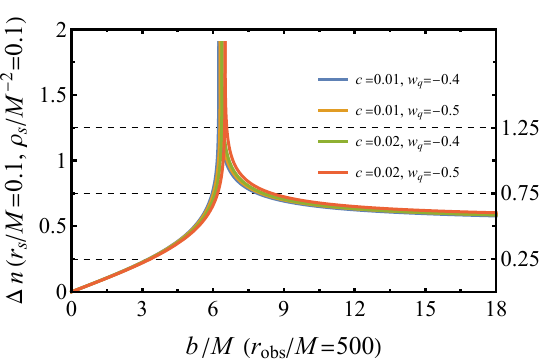}
        \includegraphics[width=0.48\textwidth]{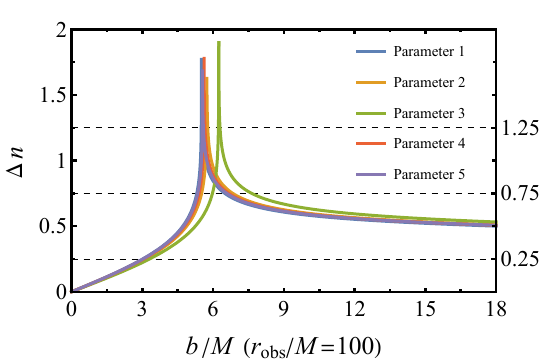}
        \hspace{0.2cm}
        \includegraphics[width=0.48\textwidth]{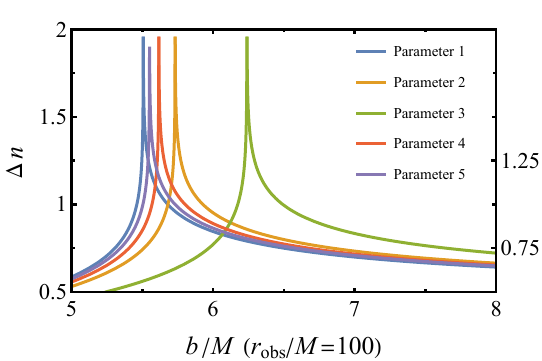}
    \caption{Variation of the total number of photon orbits $\Delta n$ with impact parameter $b$ for different spacetime parameters and observers at different positions.}
    \label{n}
\end{figure*}

Figure~\ref{n} shows the variation of the total number of photon orbits $\Delta n$ with the impact parameter $b$ for different spacetime parameters and observers at different positions. The first row illustrates the influence of observers at different positions and different DM parameters $\rho_s$ and $r_s$ on $\Delta n$ under the QF. It can be seen that as $\rho_s$ and $r_s$ increase, the overall peak shifts to the right; the ranges of the impact parameter $b$ corresponding to the lensed ring and photon ring increase. Larger $\rho_s$ (or $r_s$) also makes $\Delta n$ more sensitive to $r_s$ (or $\rho_s$). Unlike the integrated intensity, the observer’s position has  a negligible influence on $\Delta n$. The second row presents the influence of observers at different positions and different QF parameters $c$ and $w_q$ on $\Delta n$ in the Dehnen-(1,4,5/2) DM. As $c$ and $w_q$ increase, the overall peak also shifts to the right, and the ranges of the impact parameter $b$ corresponding to the lensed ring and photon ring increase, though the effect is relatively weak. 
Similarly, the observer’s position has little influence on $\Delta n$.
The third row presents the effects of different parameters in table~\ref{canshu} on $\Delta n$. 
It is clear that increases in $\rho_s$, $r_s$, $c$, and $|w_q|$ all broaden the range of the impact parameter $b$ spanned by the lensed ring and the photon ring.

\begin{figure*}[htp]
    \centering
        \includegraphics[width=0.32\textwidth]{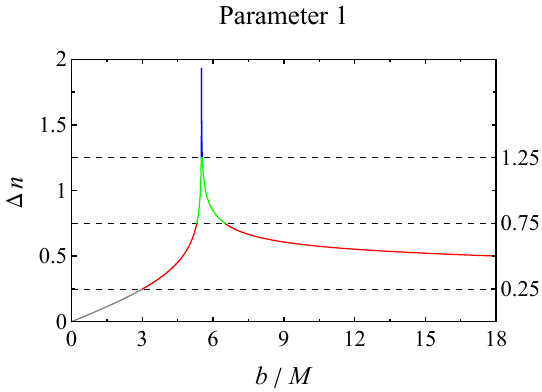}
        \includegraphics[width=0.32\textwidth]{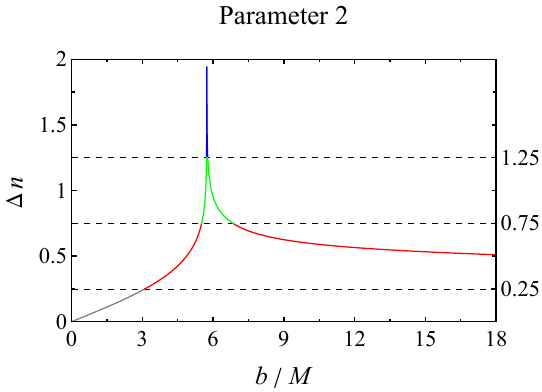}
        \includegraphics[width=0.32\textwidth]{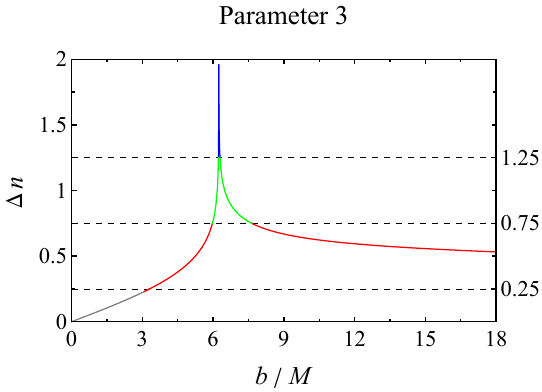}\\[1em]
        \includegraphics[width=0.32\textwidth]{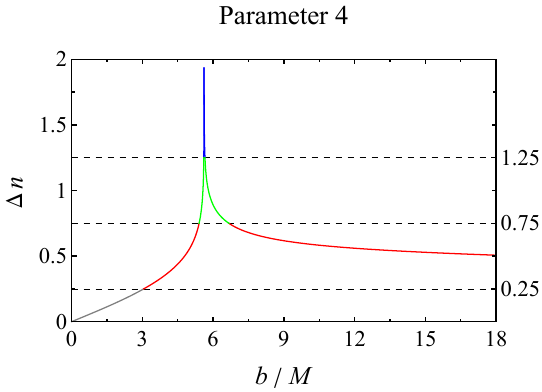}
        \includegraphics[width=0.32\textwidth]{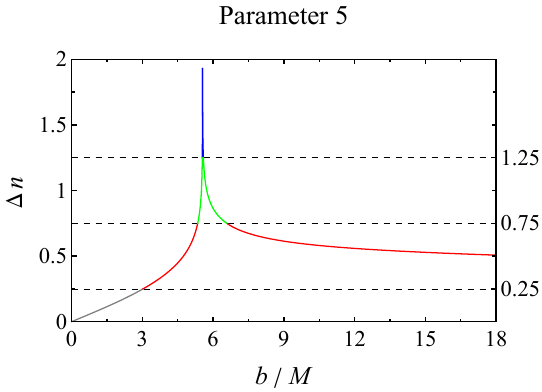}
    \caption{Variation of the total photon orbit number $n$ with the impact parameter $b$ for different parameter sets (1–5 in table~\ref{canshu}) and observer positions.}
    \label{nn}
\end{figure*}

\begin{figure*}[htp]
    \centering
        \includegraphics[width=0.32\textwidth]{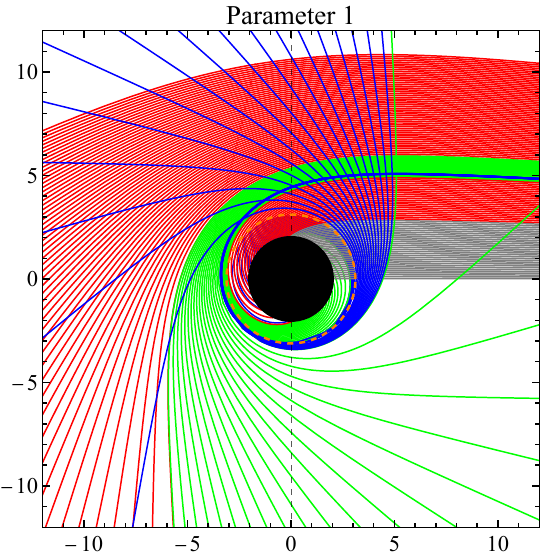}
        \includegraphics[width=0.32\textwidth]{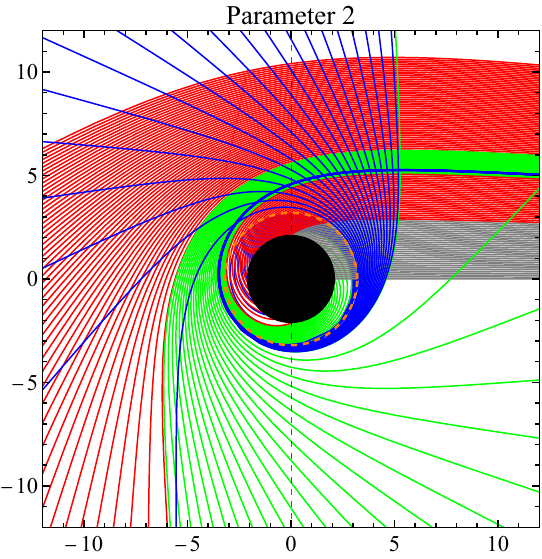}
        \includegraphics[width=0.32\textwidth]{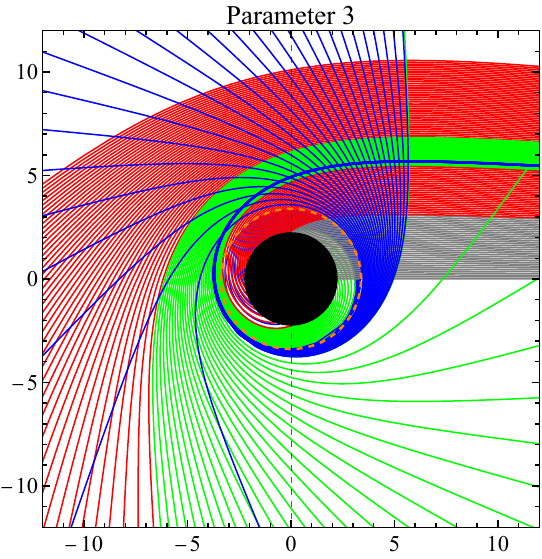}\\[1em]
        \includegraphics[width=0.32\textwidth]{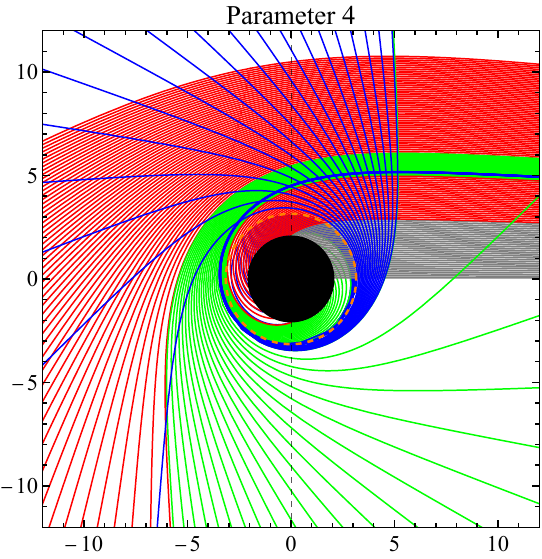}
        \includegraphics[width=0.32\textwidth]{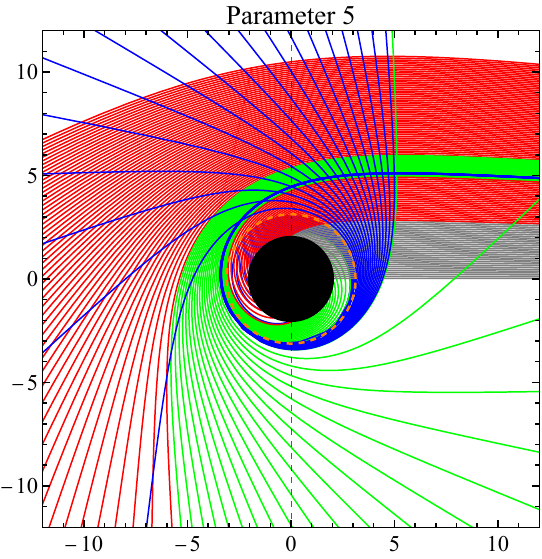}
    \caption{Trajectories of light rays for the different emission types in the spacetime configurations corresponding to those in figure~\ref{nn}.
    The red, green, and blue lines correspond to direct emission, lensed ring emission, and photon ring emission, respectively. The gray curves represent light rays not intersecting the accretion disk. This figure refines figure~\ref{guiji}. The orange dashed line and the black disk represent the photon sphere and the BH, respectively. All coordinates are scaled in units of the BH mass $M$.}
    \label{guiji2}
\end{figure*}

Figure~\ref{nn} plots the dependence of $\Delta n$ on the impact parameter $b$ for parameter sets 1–5 listed in table~\ref{canshu}.
Under different parameter sets, the trajectories of light rays corresponding to different emission types are shown in figure~\ref{guiji2}, where the observer is located at $r_\mathrm{obs}/M=100$. Rays of different colors correspond to different emission types. It should be emphasized that figure~\ref{guiji2} provides a more detailed elaboration of figure~\ref{guiji}. The specific features of each emission type visible here align consistently with the results shown in figure~\ref{nn}.

\subsection{Integral intensity and transfer function}
Assuming the radiation emitted by the stationary accretion flow on the accretion disk has a specific intensity $I_e(r)$ and frequency $\nu_e$, according to Liouville's theorem, $I_e(r)/\nu_e^3$ remains constant along the light propagation path, satisfying $I_e(r)/\nu_e^3=I_o(r)/\nu_o^3$, where $I_o(r)$ and $\nu_o$ are the specific intensity and frequency observed by the observer, respectively. Using eq.~\eqref{g} and the redshift factor eq.~\eqref{jhy} for stationary accretion, the relation between $I_o(r)$ and $I_e(r)$ is derived as
\begin{equation}
    I_o(r)=I_e(r)\dfrac{\nu_o^3}{\nu_e^3}=\sqrt{\dfrac{f^3(r)}{f^3(r_\mathrm{obs})}}I_e(r).
\end{equation}
Then, using eq.~\eqref{fobs}, the integral intensity is provided
\begin{equation}
    F_\mathrm{obs}=\int_0^\infty I_o(r)\mathrm{d}\nu_o=\dfrac{f^2(r)}{f^2(r_\mathrm{obs})}I_\mathrm{emit}(r),
\end{equation}
where 
\begin{equation}
    I_\mathrm{emit}(r):=\int_0^\infty I_e(r)\mathrm{d}\nu_e,
\end{equation}
denotes the intensity distribution emitted by the accretion disk.

\begin{figure*}[htp]
    \centering
        \includegraphics[width=0.32\textwidth]{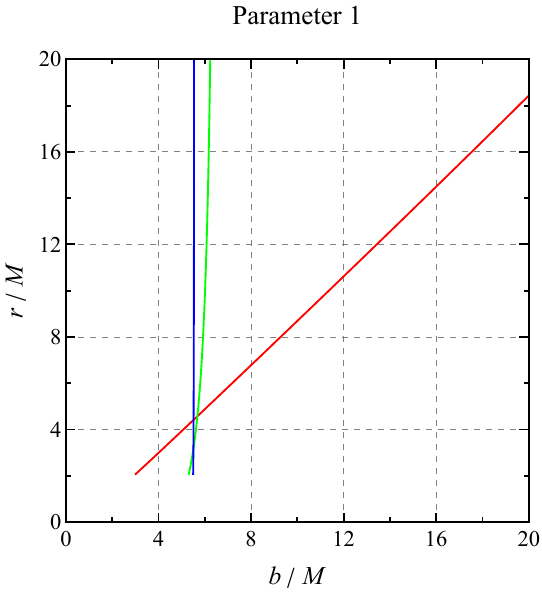}
        \includegraphics[width=0.32\textwidth]{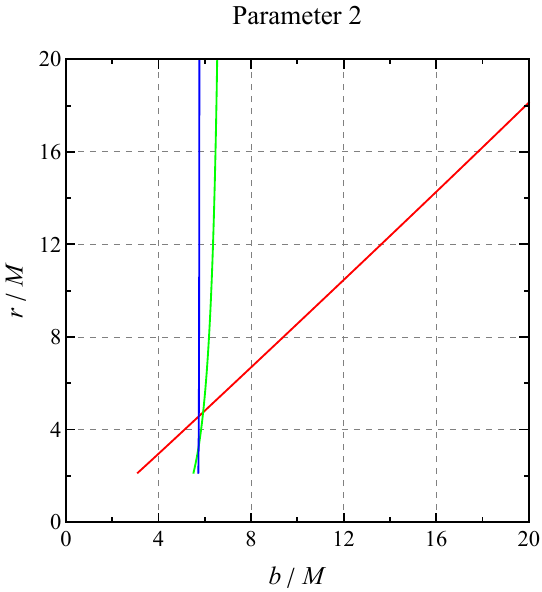}
        \includegraphics[width=0.32\textwidth]{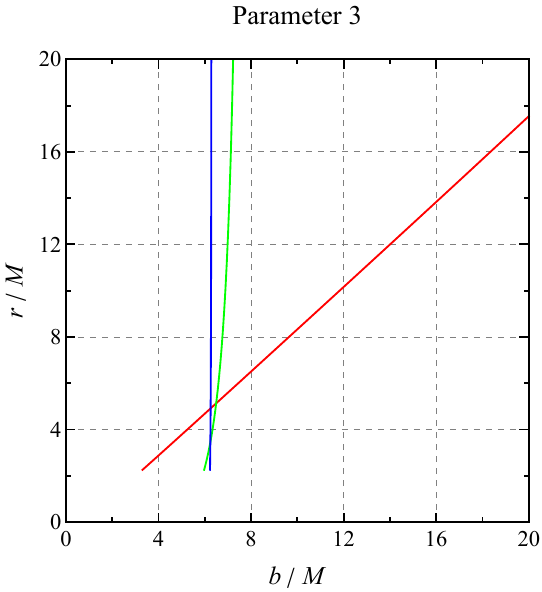}\\[1em]
        \includegraphics[width=0.32\textwidth]{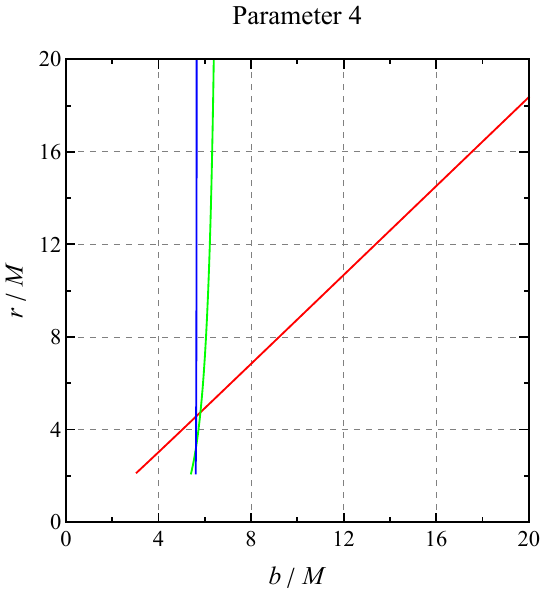}
        \includegraphics[width=0.32\textwidth]{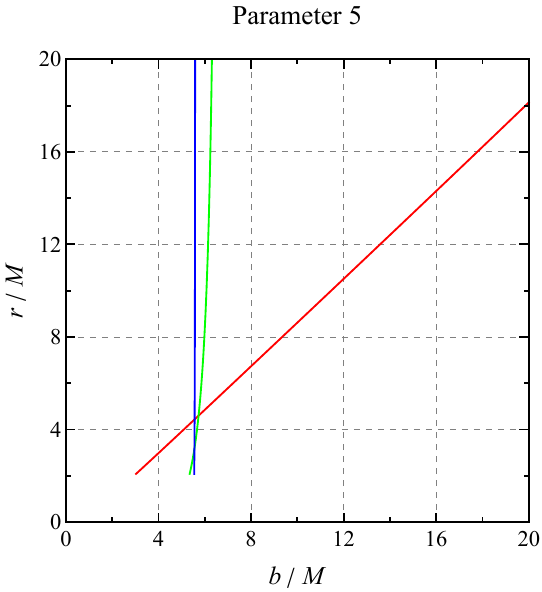}
    \caption{The first three orders of the transfer function for spacetimes with different parameters. The red, green, and blue solid lines correspond to the transfer functions for direct emission, lensed ring emission, and photon ring emission, respectively.}
    \label{chuanditu}
\end{figure*}

For an optically thin accretion disk, each intersection of a light ray with the accretion disk produces additional specific intensity. Therefore, the integrated intensity should be reformulated as the sum of the extra intensity acquired each time the light rays crosses the accretion disk, i.e.,
\begin{equation}
    F_\mathrm{obs}(b)=\sum_k\dfrac{f^2(r_k(b))}{f^2(r_\mathrm{obs})}I_\mathrm{emit}(r_k(b)),\label{pf}
\end{equation}
where $r_k(b)$ is the transfer function~\citep{r54}, representing the relation between the photon's impact parameter $b$ and the radial coordinate $r$ of the $k$-th intersection of the light rays with the accretion disk. This function can be derived from the result of eq.~\eqref{xinvarphi}, namely
\begin{equation}
    r_k(b)=\dfrac{1}{u(\pi/2+(k-1)\pi,b)},
\end{equation}
where $k=1,2,3$ correspond to the direct emission, lensed ring emission, and photon ring emission, respectively. Figure~\ref{chuanditu} shows the first three orders of the transfer function for spacetimes with different parameters, corresponding one-to-one with figure~\ref{nn}. The slope of the transfer function, $\mathrm{d}r_k(b)/\mathrm{d}b=\mathnormal{\Gamma}$, is the magnification factor. In the figure, the red curves represent the $k=1$ transfer function, corresponding to the direct emission of the accretion disk. Evidently, the transfer functions for spacetimes with different parameters all exhibit an approximately linear relation with the impact parameter $b$, i.e., $\mathnormal{\Gamma}\approx1$, indicating that the direct emission corresponds to the radiation profile of the gravitational redshift source. Indeed, figure~\ref{chuanditu} shows that direct emission constitutes the main component of the integrated intensity, as it occupies the largest proportion of the total radiation. 
The green curves represent the $k=2$ transfer function, corresponding to lensed ring emission. It should be emphasized that although the inclusion of Dehnen-(1,4,5/2) DM and QF enlarges the lensed ring, its contribution remains very small compared to direct emission.
Finally, the blue curves near $b=b_p$ are the $k=3$ transfer functions, corresponding to the photon ring emission.
The slope of the third-order transfer function approaches infinity, meaning the magnification of the photon ring imaging is infinitely reduced, and its radiative flux is practically negligible.
The boundary of the BH shadow for different parameters is determined by direct emission, a conclusion that is in principle valid for disk accretion, yet only holds for models with an overly large inner edge of the accretion disk when the contributions from the $k=2$ and $k=3$ higher-order images are neglected.
It is clear that the radiation flux decreases rapidly as $k$ increases. Thus, only the first three orders of the transfer function need to be considered in the present work, since the contribution from higher-order emissions with $k>3$ is negligible.

\subsection{Observational signatures of BHs embedded in Dehnen-(1,4,5/2)-type DM halo with a QF}
In the thin-disk model, the accretion disk serves as the primary radiation source illuminating the BH. The emission intensity distribution of the disk, $I_\mathrm{emit}(r)$, depends not only on the radiation profile but also on the location of the disk’s inner edge $r_\mathrm{in}$. This subsection analyzes the image features of BHs embedded in a Dehnen-(1,4,5/2)-type DM halo with a QF, considering three distinct emission profiles of the accretion disk. It should be emphasized that, for simplicity, a monochromatic radiation model is still adopted.

\begin{figure*}[htp]  
    \centering  
        \includegraphics[width=0.32\textwidth]{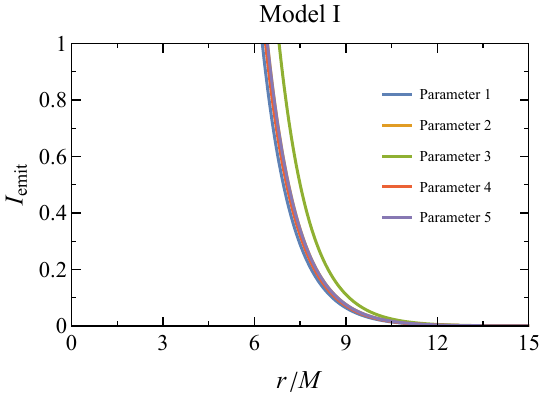}  
        \includegraphics[width=0.32\textwidth]{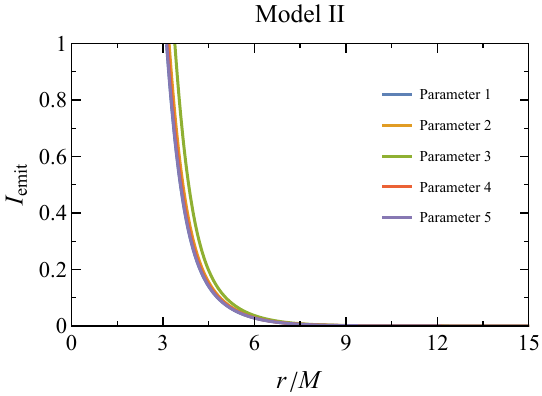}  
        \includegraphics[width=0.32\textwidth]{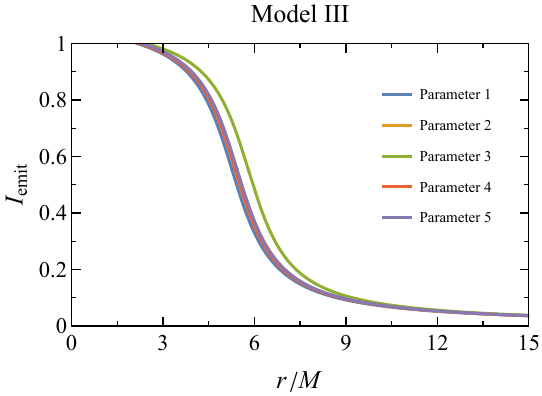}  
    \caption{The radiation intensity distribution $I_\mathrm{emit}$ of the accretion disk as a function of $r$ is shown for three different accretion disk models with parameter sets listed in table~\ref{canshu}.}  
    \label{pan}  
\end{figure*}

The emission intensity distribution of the disk for the three models we consider is given respectively by:  
\begin{itemize}  
    \item Model I  
\end{itemize}  
\begin{equation}  
    I_\mathrm{emit}(r)=\begin{cases}  
        \exp{(-r+r_\mathrm{ISCO})},\quad &r>r_\mathrm{ISCO},\\[1em]  
        0,\quad &r\le r_\mathrm{ISCO}.  
    \end{cases}  
\end{equation}  
\begin{itemize}  
    \item Model II  
\end{itemize}  
\begin{equation}  
    I_\mathrm{emit}(r)=\begin{cases}  
        \dfrac{2-\tanh{(r-r_p)}}{2}\exp{(-r+r_p)},\quad &r>r_p,\\[1em]  
        0,\quad &r\leq r_p.  
    \end{cases}  
\end{equation}  
\begin{itemize}  
    \item Model III  
\end{itemize}  
\begin{equation}  
    I_\mathrm{emit}(r)=\begin{cases}  
        \dfrac{\pi/2-\arctan{(r-r_\mathrm{ISCO}+1)}}{\pi/2+\arctan{(r_p)}},\quad &r>r_h,\\[1em]  
        0,\quad &r\leq r_h.  
    \end{cases}  
\end{equation}  
For the three models discussed above, the inner edge of the accretion disk is set at three characteristic radii respectively: the ISCO radius of massive particles $r_\mathrm{in}=r_\mathrm{ISCO}$, the photon sphere radius $r_p$, and the event horizon radius $r_h$.
The ISCO radius $r_\mathrm{ISCO}$ is given by  
\begin{equation}  
    r_\mathrm{ISCO}=\dfrac{3f(r_\mathrm{ISCO})f'(r_\mathrm{ISCO})}{2f'^2(r_\mathrm{ISCO})-f(r_\mathrm{ISCO})f''(r_\mathrm{ISCO})},  
\end{equation}  
where $f'(r)$ and $f''(r)$ denote the first and second derivatives of the function $f(r)$ with respect to $r$, respectively. Figure~\ref{pan} shows the distributions of $I_\mathrm{emit}(r)$ for the five different parameter sets listed in table~\ref{canshu} under the different models. It can be seen that increasing the DM parameters ($\rho_s$ and $r_s$) and QF parameters ($c$ and $w_q$) not only enlarges the event horizon and photon sphere radii, but also increases the ISCO radius.
Among the three accretion disk models, the second exhibits the most rapid variation in $I_\mathrm{emit}$, while the third varies the most slowly.

\begin{figure*}[htp]
    \centering
        \includegraphics[width=0.48\textwidth]{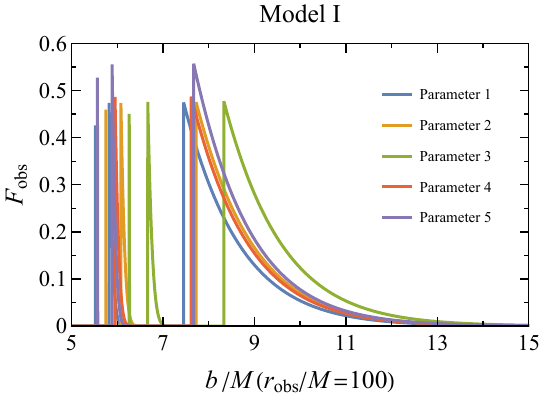}
        \hspace{0.2cm}
        \includegraphics[width=0.48\textwidth]{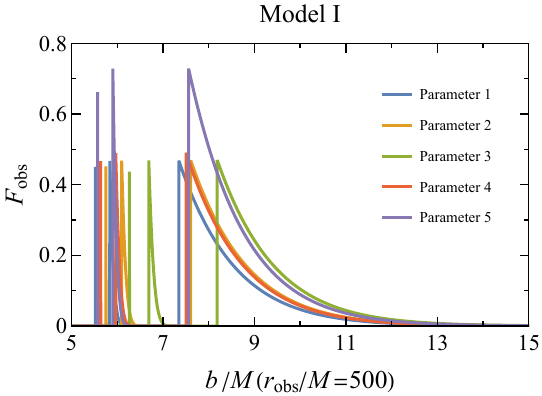}
        \includegraphics[width=0.48\textwidth]{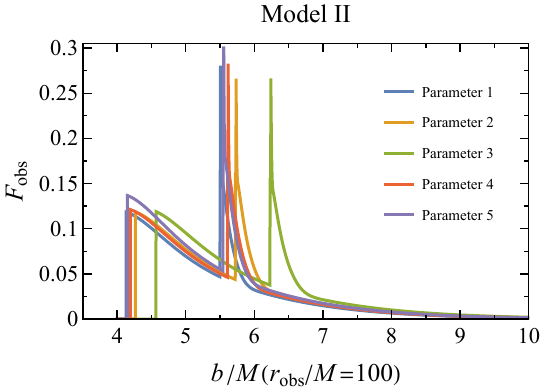}
        \hspace{0.2cm}
        \includegraphics[width=0.48\textwidth]{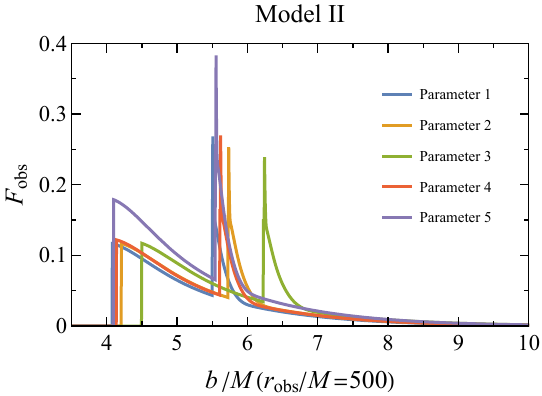}
        \includegraphics[width=0.48\textwidth]{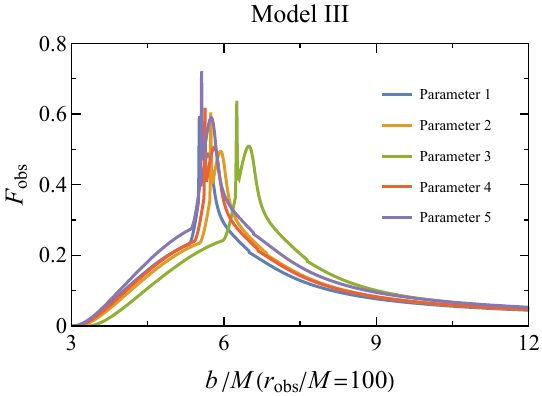}
        \hspace{0.2cm}
        \includegraphics[width=0.48\textwidth]{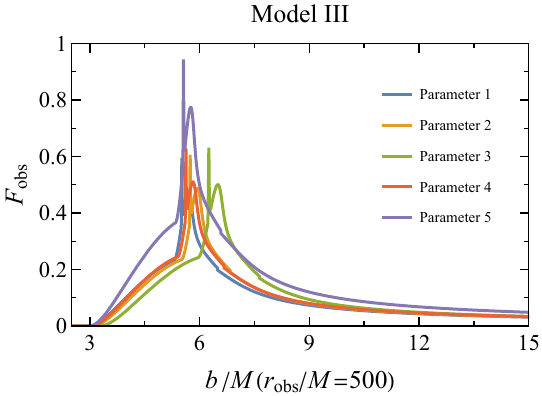}
    \caption{Integrated intensity $F_\mathrm{obs}$ distributions for three accretion disk models as seen by observers at different positions in spacetimes with various parameters (1–5 in table~\ref{canshu}).}
    \label{panobs}
\end{figure*}

According to eq.~(\ref{pf}), we plot the integrated intensity distributions for three accretion disk models as seen by observers at different positions in spacetimes with various parameter sets (1–5 in table~\ref{canshu}) in figure~\ref{panobs}. 
The first row displays the integrated intensity distributions of Model I for observers with different parameter sets located at $r_\mathrm{obs}/M=100$ (left) and $r_\mathrm{obs}/M=500$ (right).
It can be clearly seen that $F_\mathrm{obs}$ overall exhibits three peaks, corresponding from left to right to the photon ring, the lensed ring, and the direct emission.
The photon ring and lensed ring are separated from the direct emission, and their positions match those in figure~\ref{nn} and figure~\ref{chuanditu}. 
The second row shows the integrated intensity distributions of Model II for observers with different parameter sets (1–5 in table~\ref{canshu}) at different locations.
Unlike Model I, here the photon ring and lensed ring are directly superimposed on the direct emission.
$F_\mathrm{obs}$ shows two peaks from left to right: the first peak corresponds to the direct emission, and the second peak results from all three emission mechanisms.
Finally, the third row presents the integrated intensity distributions of Model III for observers with different parameter sets (1–5 in table~\ref{canshu}) at different locations. 
$F_\mathrm{obs}$ exhibits two closely spaced peaks, the first of which is dominated by emission from the photon ring, while the second arises predominantly from the lensed ring contribution.

In all three models, the effects of various spacetime parameters on the observed integrated intensity $F_\mathrm{obs}$ are consistent with the spherical accretion case.
Specifically, the influences of $\rho_s$ and $r_s$ on $F_\mathrm{obs}$ are almost independent of the observer’s position.
The effect of the QF parameter $c$ on $F_\mathrm{obs}$ exhibits only a weak dependence on the observer’s position at small values of $|w_q|$, which is fully consistent with the results presented in section~\ref{sub4} (see the second rows of figures~\ref{fobsxx} and~\ref{fobsj}).
In sharp contrast, the influence of $w_q$ on $F_\mathrm{obs}$ depends strongly on the observer’s position: for a fixed $w_q$, a more distant observer yields a higher integrated intensity, and larger values of $|w_q|$ further enhance this effect. These results provide crucial insights for interpreting the image features arising from the interaction of Schwarzschild BHs with DM and DE in specific accretion environments, as well as for constraining the equation-of-state parameter $w_q=p_q/\rho_q$.

\begin{figure*}[htp]
    \centering
        \includegraphics[width=0.44\textwidth]{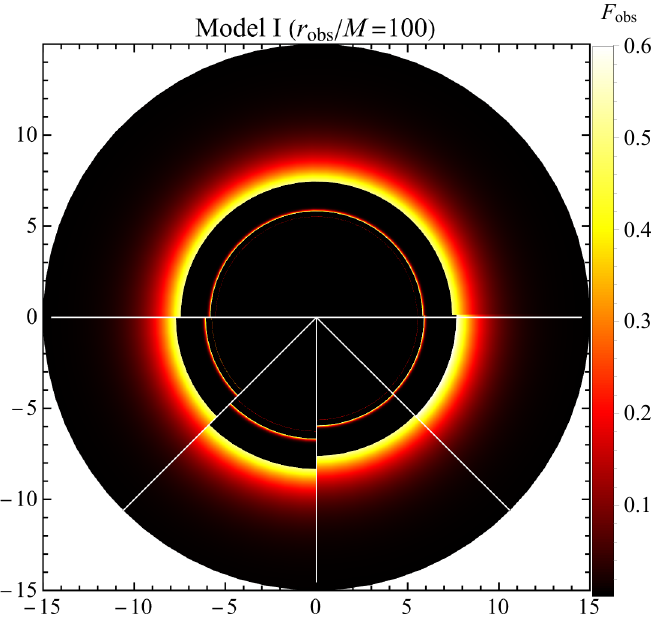}
        \hspace{0.2cm}
        \includegraphics[width=0.44\textwidth]{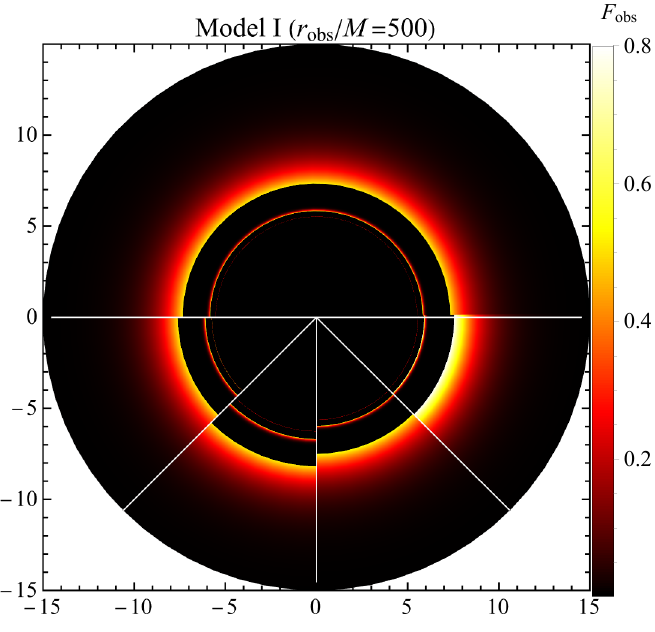}
        \includegraphics[width=0.44\textwidth]{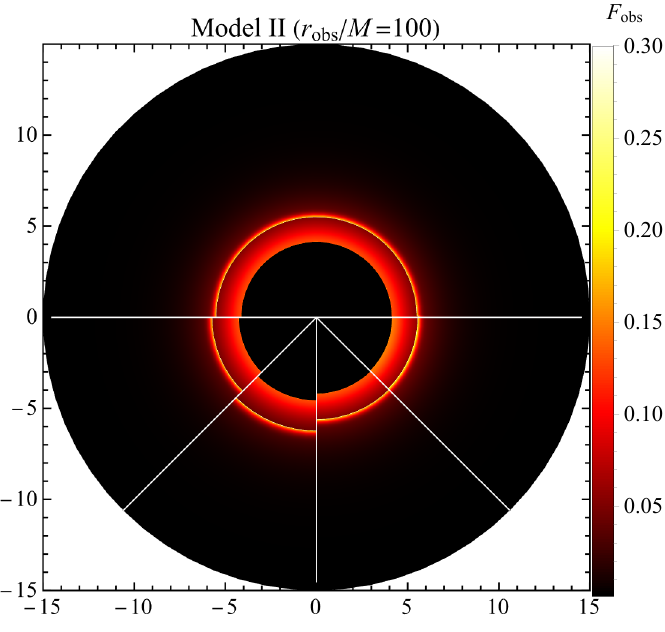}
        \hspace{0.2cm}
        \includegraphics[width=0.44\textwidth]{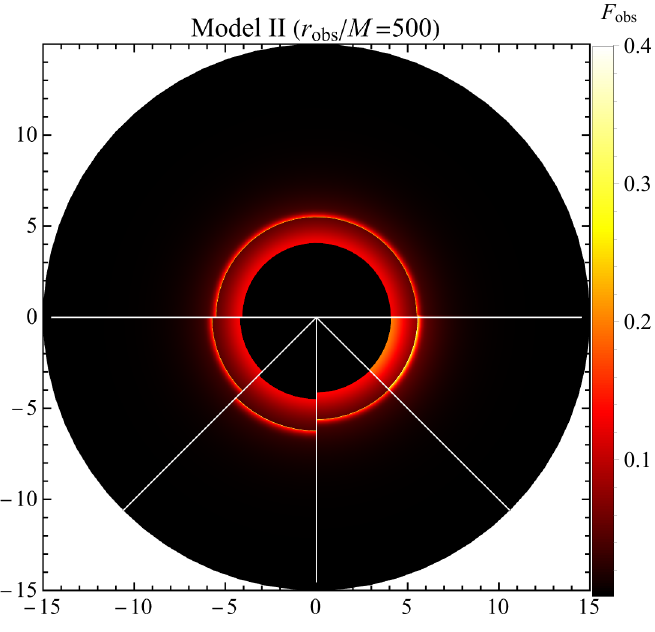}
        \includegraphics[width=0.44\textwidth]{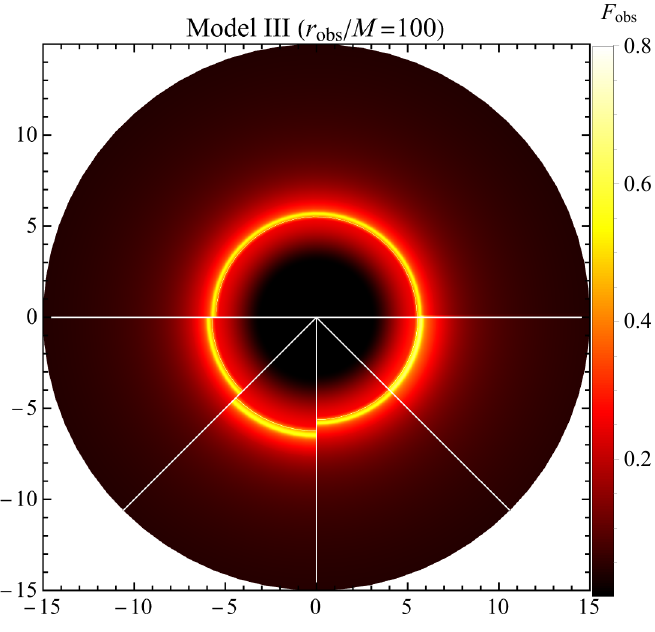}
        \hspace{0.2cm}
        \includegraphics[width=0.44\textwidth]{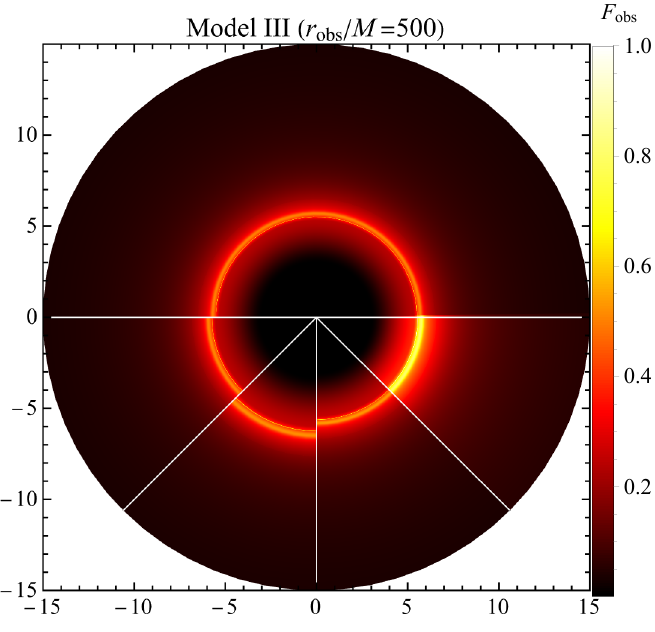}
    \caption{BH images are plotted based on figure~\ref{panobs}. The panels are arranged by polar angle ranges: the panel covering $0$–$\pi$ corresponds to parameter 1 in table \ref{canshu}; the panel covering $\pi$–$5\pi/4$ corresponds to parameter 2; the panel covering $5\pi/4$–$3\pi/2$ corresponds to parameter 3; the panel covering $3\pi/2$–$7\pi/4$ corresponds to parameter 4; and the panel covering $7\pi/4$–$2\pi$ corresponds to parameter 5.}
    \label{pbh}
\end{figure*}

We plot the corresponding BH images based on figure~\ref{panobs} in figure~\ref{pbh}. It can be intuitively seen that, unlike spherical accretion, the shadow size of the disk accretion model depends on the accretion disk model. 
Concretely, Model I corresponds to the largest shadow, followed by Model II, and Model III is the smallest. 
Larger $|w_q|$ values and more distant observers yield higher intensity (the fifth panel of the second column in figure~\ref{pbh}).
The contribution to $F_\mathrm{obs}$ mainly comes from direct emission, while the lensed ring and photon ring occupy too small a region compared to direct emission, resulting in a minor contribution to $F_\mathrm{obs}$.

Finally, we briefly discuss the observability of the aforementioned ring structures and the effects of DM and DE. The current angular resolution of the EHT is sufficient to measure the overall shadow size of M87* and Sgr A*, but cannot resolve the fine structures of the photon ring and lensing ring, whose typical widths are only a few microarcseconds. The ring broadening and radial displacement induced by the Dehnen-(1,4,5/2) DM halo and QF in this work also lie below the current resolution limit, and thus cannot be directly tested with existing EHT data.

Looking ahead, by increasing the number of telescopes and extending interferometric baselines to significantly improve the angular resolution to approximately $4~\mu\mathrm{as}$, it will become possible to resolve the lensing ring substructure with the enhanced resolution, while resolving the photon ring will require even higher angular resolution. At that point, the diagnostic method proposed in this work can be applied to impose observational constraints on the DM and DE environments around supermassive BHs via their fine structures.

\section{Conclusions and discussion\label{sub6}}
In this work, we systematically investigate the optical properties and shadow imaging features of static BHs embedded in a Dehnen-(1,4,5/2) DM halo with a QF. Following a self-consistent framework, we first characterize the background spacetime structure and photon geodesic properties, then explore the BH observational signatures under two canonical accretion frameworks, namely spherical accretion and thin-disk accretion, and finally derive robust observational discriminants for the effects of DM and DE.

Before proceeding to the BH imaging, we explore the foundational geometric structure governing photon propagation in the composite DM-DE spacetime. Specifically, we first review the metric of the background spacetime and numerically solve for the event horizon radius under different parameter sets. Subsequently, the null geodesic equations for photons are derived, the photon effective potential is constructed, and the critical physical quantities—including the photon sphere radius and critical impact parameter—are calculated. To visualize these results, we present the variation of the event horizon with DM and QF parameters, the effective potential curves for different spacetime configurations, and the corresponding light ray trajectories using a ray-tracing algorithm. These numerical calculations reveal the distinctive geometric properties of the composite DM–DE spacetime. Increases in either the DM halo parameters ($\rho_s$, $r_s$) or the QF parameters ($c$, $|w_q|$) lead to a systematic enlargement of all critical physical scales, including the event horizon radius, photon sphere radius, critical impact parameter, and shadow radius. Notably, within a specific parameter range (e.g., $c<1/8M$, $w_q=-2/3$), the QF induces an additional cosmological horizon outside the BH event horizon, which confines the observer to a finite domain of outer communication bounded by these two horizons.

As the first core objective of this work, the optical characteristics of the BH under the spherical accretion framework are systematically studied. We derive general analytical expressions for the redshift factor and observed integrated intensity under both infalling and static spherical accretion models, and rigorously show that static accretion corresponds to the limiting case of infalling accretion in which the radial velocity of the accretion flow vanishes everywhere. To visualize these analytical results and facilitate a direct comparison between the two accretion models, we select five representative sets of spacetime parameters (to probe the individual effects of increasing $\rho_s$, $r_s$, $c$ and $|w_q|$), plot the integrated intensity profiles, and generate the BH images for each configuration. Our key findings are summarized below.
\begin{itemize}
    \item \textit{General properties of spherical accretion}: The BH shadow radius is identical under both accretion modes and equals the critical impact parameter, confirming that the shadow size is a purely geometric quantity determined solely by the background spacetime. The shadow intensity under infalling accretion is consistently lower than that under static accretion, a direct consequence of the Doppler redshift induced by the inward radial motion of the infalling flow.
    \item \textit{Key DM and DE effects}: The DM and QF parameters exert diametrically opposite and observationally distinguishable effects on BH shadow features: increases in the DM parameters $\rho_s$ and  $r_s$  enlarge the shadow radius but reduce the shadow intensity, with their effects nearly independent of the observer’s position; in contrast, increases in the QF parameters $c$ and $|w_q|$ significantly enhance the shadow intensity but have negligible influence on the shadow radius, with the intensity modulation effect exhibiting a strong dependence on the observer’s distance (a more distant observer measures a higher intensity for fixed increases in QF parameters). We further uncover a mutual enhancement of parameter sensitivity: a larger value $\rho_s$ (or  $r_s$) amplifies the sensitivity of the intensity to variations in $r_s$  (or $\rho_s$), and an identical cross-sensitivity enhancement exists between the QF parameters $c$ and $|w_q|$.
\end{itemize}

As the second core objective of this work, the optical characteristics of the BH under the more astrophysically realistic thin-disk accretion framework are further studied. To elucidate the underlying gravitational lensing mechanisms, we begin by classifying the light rays emitted from the disk into three physically distinct categories based on the total number of orbits around the BH: direct emission rays, lensed ring rays, and higher-order photon ring rays. We then plot the deflection angle as a function of the impact parameter (using the same parameter sets as the spherical accretion analysis), construct refined light ray trajectories based on the above classification, and generate the corresponding profile of the transfer functions. 
Three distinct thin accretion disk models, differentiated by their inner radiative boundary (set at the ISCO, the photon sphere, and the event horizon, respectively), are systematically examined. To compare their observational signatures, we plot the radial emission profiles, compute the integrated intensity distributions using our derived analytical formulas, and generate the corresponding BH images for each model.
Our main conclusions are summarized below.
\begin{itemize}
    \item \textit{General properties of thin-disk accretion}: The observed integrated intensity is overwhelmingly dominated by direct emission across all three models. The contributions from the lensed ring and photon ring are negligible owing to their extremely narrow spatial width. The radial positions of the lensed ring and photon ring remain almost independent of the observer’s location, providing a stable geometric probe of the spacetime.
    \item \textit{Key DM and DE effects}: Unlike the spherical accretion case accretion case, the apparent shadow radius in the thin-disk accretion is governed by the choice of the disk’s inner boundary. The model with the ISCO as the inner boundary yields the largest shadow, followed by the photon sphere model, while the event horizon model produces the smallest shadow. The inner boundary also determines the separability of the three emission components. When the ISCO is adopted, the direct, lensed‑ring, and photon‑ring emissions are clearly separated. In contrast, when the inner boundary is set to the photon sphere or event horizon, the ring components are superimposed onto the direct emission. We further find that increases in either DM or QF parameters lead to a significant spatial broadening of the lensed and photon ring regions, with a mutual enhancement of parameter sensitivity identical to that in the spherical accretion case. Consistent with the spherical accretion results, the influence of the QF parameter $w_q$ on the integrated intensity depends strongly on the observer’s position, while the effect of the QF normalization parameter $c$ is nearly independent of the observer’s location for small values of $|w_q|$.
\end{itemize}

Furthermore, we provide a physical interpretation for the distinct influences of DM and DE on BH images, especially for the strong observer-distance dependence of the QF effect. The Dehnen-(1,4,5/2) DM halo is a localized matter distribution with positive transverse pressure, whose gravitational effect is confined to a few scale radii around the BH; as a result, its modifications to the shadow size and integrated intensity are nearly independent of the observer’s radial position. In contrast, the QF is a cosmic-scale energy component with negative pressure, which fundamentally alters the asymptotic structure of spacetime and introduces a cosmological horizon. Photons propagating from the BH to the observer thus accumulate gravitational redshift continuously along their entire path, and the integrated intensity increases accordingly with the observer’s distance. This intrinsic difference between the local DM effect and the global DE effect offers a firm physical basis for distinguishing the two components through BH imaging observations.

Taken together, the decoupled DM and DE signatures in BH images—obtained from both spherical and face-on thin-disk accretion—provide a robust observational criterion to distinguish DM-dominated from DE-dominated environments and to independently constrain the DE equation-of-state parameter $w_q=p_q/\rho_q$. To bridge the gap between our idealized modeling and real observations, several critical extensions are warranted:
\begin{itemize}
    \item \textit{Inclined accretion disks}: Existing EHT images of M87* and Sgr A* arise from inclined, rather than face-on, accretion flows. Systematic analysis of inclined thin-disk models is therefore essential for the astrophysical interpretation of these data.
    \item \textit{Rotating accretion flows}: Realistic accretion flows inevitably carry angular momentum and orbit the central compact object. Since our present analysis is restricted to a static, face-on configuration, a key open question is whether the distinct DM and DE signatures persist and remain observationally distinguishable when the accretion disk rotates around a static BH—a scenario that calls for dedicated theoretical and numerical study.
    \item  \textit{Extension to rotating BHs}: Realistic EHT targets such as M87* and Sgr A* are rotating Kerr BHs, whose frame‑dragging effect breaks spherical symmetry and produces a D‑shaped shadow asymmetry that increases with spin. When the Dehnen DM halo is superposed on the Kerr background, its gravitational field couples through curvature, enlarging the photon capture cross section and shifting the horizon and photon‑sphere radii. In contrast, the cosmic scale QF DE alters the asymptotic spacetime structure and introduces observer‑distance‑dependent effects. When combined with spin, we conjecture that the frame-dragging effect and the distance-dependent DE effect together produce a unique global coupling effect, which leads to a systematic variation of shadow asymmetry with observer distance. Since spin, DM and DE all affect shadow size and shape, significant parameter degeneracies are expected. Therefore, generalizing our static framework to rotating BHs is essential for correctly interpreting EHT observations and breaking these degeneracies, and this constitutes a core direction of our future work.
\end{itemize}
Beyond these geometric extensions, the universality of our conclusions needs to be systematically verified across a broader class of dark sector scenarios (e.g., NFW DM halos~\citep{c1,c2}, isothermal DM profiles~\citep{c3}, phantom DE models~\citep{c4}, and coupled DM-DE interaction systems~\citep{c5}) and in modified gravity frameworks~\citep{c6,c7,c8,c9,c10}. Additionally, next-generation EHT observations will resolve fine-scale lensing rings and photon rings, enabling us to break existing parameter degeneracies and improve the precision of DM and DE constraints. Thick-disk models, most notably advection-dominated accretion flows, can also be incorporated to better represent the low-luminosity supermassive BHs targeted by the EHT.

\section*{Acknowledgements}
This work is supported by the National Key R\&D Program of China (Grant No. 2024YFA1611700), the National Natural Science Foundation of China (Grants Nos. 12105039, 12133003, 12494574 and 12326602), the Guangxi Key R\&D Program (Guangxi Funeng Action Plan, Grant No. Guike FN2504240040), the Guangxi Science and Technology Innovation Platform Program (Leitai Action Plan, Grant No. Guike LT2600640026), and the Guangxi Highland of Innovation Talents Program.

\section*{Data availability statement}
This manuscript has no associated data. [Authors' comment: Data sharing is not applicable to this article as no datasets were generated or analysed during the current study.]

\section*{Code availability statement}
 Code/software will be made available on reasonable request. [Authors' comment: The code/software generated during and/or analysed during the current study is available from the corresponding author on reasonable request.]








\printcredits

\bibliographystyle{cas-model2-names}

\bibliography{cas-refs}



\end{document}